\documentclass[a4paper,11pt]{article}
\pdfoutput=1 

\usepackage{jcappub} 

\usepackage[T1]{fontenc} 

\usepackage{tikz}
\usepackage{hyperref}
\usepackage{bm}
\usetikzlibrary{arrows}

\title{\boldmath Constraining Statistical Isotropy using 21cm Power Spectrum and Bispectrum}

\def\d{\mathrm{d}}

\author[a]{Bhuwan Joshi,}
\author[a]{Rahul Kothari}

\affiliation[a]{School of Physical Sciences, Indian Institute of Technology, Mandi, Himachal Pradesh, 175005}

\emailAdd{bhuwanj230@gmail.com}
\emailAdd{quantummechanicskothari@gmail.com}

\abstract{The Cosmological Principle states that the universe is statistically isotropic and homogeneous on large length scales, typically $\gtrsim 70$Mpc. A detection of significant deviation would help us falsify the simplest models of inflation. In this regard, there are potential indications of departures from this principle, e.g., observations from WMAP and Planck show signs of a preferred direction in the temperature fluctuations known as hemispherical asymmetry in CMB. Phenomenologically, this has been studied using a dipole modulation model. In addition to this, a number of possible mechanisms have been proposed in the literature to explain this anomaly. Some of these scenarios generate dipolar asymmetry or predict quadrupolar asymmetry in the primordial power spectrum of curvature perturbations. In this paper, we study both these asymmetries. To fulfill the objective, we employ 21cm intensity mapping technique post during post-reionization era, i.e., $z\lesssim 7$. We apply Fisher formalism to constrain dipolar and quadrupolar anisotropy parameters using both 21cm power and bispectra and give forecasts for three intensity mapping surveys: SKA-Mid, HIRAX and PUMA. Although 21cm intensity mapping is a very promising cosmological probe, the signals are severely affected by foregrounds. To mitigate the foreground effects, we use foreground avoidance approach. For the interferometer mode of operation, we also include the wedge effect. From our analysis, we find that PUMA, on account of its high redshift range is able to constrain both dipolar and quadrupolar parameters to better than $\sim 10^{-3}$ for redshifts $z \gtrsim 1$. This is one order of magnitude better constraints as compared to those provided by the latest CMB surveys. We also find that as compared to power spectrum, the constraining power of bispectrum is more sensitive towards foregrounds.
}

\begin{document}
\maketitle
\flushbottom

\section{Introduction \label{sec:intro}}
The Cosmological Principle (CP hereafter), i.e. the notion of statistical isotropy and homogeneity is the fundamental notion on 
which modern cosmology is based (for a comprehensive review see \cite{Aluri:2022hzs}). A detection of considerable amount of deviation would help us falsify the simplest models of inflation, e.g., single field slow roll inflation, thereby entailing the necessity of new postulates. 

CMB dipolar asymmetry, aka Hemispherical Power Asymmetry (HPA), is one example where we find a potential violation of CP. This effect was first observed in WMAP 2003 data  at $3\sigma$ level \cite{Eriksen:2003db} and has still persisted in Planck and later WMAP data sets at $\gtrsim 3.0\sigma$ level \cite{Hoftuft:2009rq, Eriksen:2007pc, Planck:2013lks, Planck:2015igc}. The latest Planck 2018 results \cite{Planck:2019evm} are also consistent with these findings. Phenomenologically, it has been studied using a dipole modulation model \cite{Gordon:2005ai, Gordon:2006ag, Prunet:2004zy, Bennett:2011}. Surprisingly, though, the quasar distribution  contradicts this observation \cite{Hirata:2009ar} showing no asymmetry.

This power asymmetry can be modelled as a spatial modulation of primordial power spectrum. It is thus of interest to seek  models of inflation that can produce the required difference in the 3D power that explains HPA while keeping the quasar bounds fixed. In the literature, a number of possible mechanisms have been put forth, including cosmological effects, tensor-mode contribution, spacetime non-commutativity \cite{Kothari:2015xva}, etc. Some scenarios (a) simultaneously generate dipolar power asymmetry and non-Gaussianity \cite{Schmidt:2012ky, Lyth:2013vha, Abolhasani:2013vaa} or (b) predict a quadrupolar power asymmetry \cite{Bartolo:2012sd}.

In the present work, our objective is to study both dipolar and quadrupolar asymmetries in the primordial power spectra using 21cm intensity mapping during post reionization era, i.e., $z\lesssim 7$. We put constraints on the anisotropy parameters using power  and bispectrum employing Fisher forecasts for intensity mapping surveys: SKA-Mid \cite{SKA}, HIRAX \cite{Crichton:2021hlc, Newburgh:2016mwi} and PUMA \cite{CosmicVisions21cm:2018rfq}. To our knowledge, only two such studies have hitherto been done in the literature. In \cite{Shiraishi:2016omb}, Shiraishi et. al, tested the statistical isotropy violation using 21cm signals during the cosmic dark ages where the effect of foreground isn't that prominent. In their work, they focused only on the 21cm power spectrum and gave forecast on both dipolar and quadrupolar anisotropy parameters for the futuristic radio telescope SKA. They found $1\sigma$ error on dipolar modulation amplitude is $\lesssim10^{-3}$ compared to $\sim 10^{-3}$ for an ideal  CMB experiment. For quadrupolar anisotropy, the constraints were $\lesssim10^{-3}$. 
In another study done by Li et al. \cite{Li:2019bsg}, authors tested the dipolar anisotropy from the epoch of reionization for SKA and Omniscope. 
They constrained the dipolar anisotropy parameter using the 21cm power spectrum while accounting for foregrounds using foreground avoidance methods, which rely on the fact that spectrally smooth foregrounds occupy a well-defined region of $k$-space, typically at the smallest $k$ modes along the line of sight $k_\|$. We thus remove all radial modes $k_\|$ smaller than $k_{||,\mathrm{min}}$ as these won't be distinguishable from the foregrounds. 
They found that SKA Phase 2 is capable of constraining the dipolar modulation amplitude parameter $\sim 10^{-2}$ for Fourier modes that range between $0.056$ and $0.15\ \mathrm {Mpc}^{-1}$. They also found that the constraints on the parameter can be improved by an order of magnitude using Omniscope.  

In the present paper, we supplement all these prior studies by including bispectrum in the analysis and also foreground effects which play a significant role in the post reionization era. The foreground effects are incorporated by employing a technique called foreground avoidance. We perform Fisher analysis for power spectrum and bispectrum to estimate the constraining power of 21cm surveys.  We find that the cumulative errors obtained on {both dipolar and} quadrupolar anisotropies 
is $\lesssim 10^{-3}$ which comparing with the CMB Planck results was of the order $\lesssim 10^{-2}$ \cite{Planck:2015inflation,Planck:2018jri}. Thus, next generation 21cm surveys would be able to provide one order of magnitude improvement in the constraining power {with respect to CMB}. We also find that the constraints using bispectrum are more sensitive to foregrounds as compared to power spectrum. This means that percentage  change in the cumulative error when changing the foreground cut is more in case of bispectrum as compared to power spectrum.  Our analysis suggests that both power and bispectrum have almost the same constraining power. We also find that PUMA works the best of all of the surveys used in this study. 



{The paper is structured in the following manner. In \S \ref{sec:21cm_Basics}, we start with a brief review of the physics of the 21cm line including a discussion of (a)  various survey modes of 21cm or HI intensity mapping and (b) foreground problem and how it is avoided. This is followed by a brief discussion of primordial asymmetries considered in this paper in \S \ref{sec:AsymmTypes}. After this, in \S \ref{sec:Power_Bispec}, we present  the calculation of power spectrum and bispectrum in the presence of asymmetries. The Fisher forecasts results are presented in \S \ref{sec:res_method}. The effects of non-linearity decrease as $z$ becomes larger. We find relatively larger contribution to the Fisher matrix at large $z$ values. Thus, for the purpose of this paper, we restrict ourselves to the linear regime. Also,  while making forecasts, we consider all relevant effects like redshift space distortions, fingers of God, instrumental noise, effect of telescope beam, etc.  We conclude in \S \ref{eq:Conclu_Outlook}. In this paper, we use fiducial $\Lambda$CDM cosmology with the best fit parameters from Planck 2018 paper \cite{Planck:2018vyg}: $h$ = 0.6766, $\Omega_\mathrm{b}h^2 = 0.02242 $ , $\Omega_\mathrm{c}h^2$ = 0.11933, $n_\mathrm{s}$ = 0.9665, $A_\mathrm{s} = 2.142\times 10^{-9}$.

\section{21cm Cosmology \label{sec:21cm_Basics}}
In this era of precision cosmology, cosmological models can be tested against the data using CMB and Large-scale structures (LSS). In addition to the next-generation surveys of galaxy number counts, we also have intensity mapping surveys that will target the integrated 21cm spectral line emission of neutral hydrogen, the most abundant element in the universe. These surveys  are advantageous as they don't attempt to resolve individual galaxies and therefore they can cover large volumes rapidly. One also obtains extremely accurate redshifts directly from imaging. The 21cm signal is an incredibly useful probe to study the Universe, e.g., cosmic dark ages and cosmic dawn can be explored with the help of 21cm signals \cite{Ali-Haimoud:2013hpa,Mondal:2023xjx,Saurabh_RRI:2021mxo},  to study the dark matter Physics, testing general relativity, etc., \cite{Darkmatter_21cm:2023slb,Darkmatter_21cm_2:2016sur, {Hall:2012wd}}.

\subsection{A brief review of the Physics of 21cm Line}
The 21cm line or HI line corresponds to the hyperfine transition between two closely spaced energy levels of the hydrogen atom. These energy levels are known as \textit{singlet} and \textit{triplet} states.  
The relative abundance of the singlet ($n_0$) and triplet states ($n_1$) can be written as \cite{21cm_BOOK, 21cm_review},
\begin{equation}
    \frac{n_1}{n_0}=\frac{g_1}{g_0}\exp{\left(-\frac{T_*}{T_S}\right)} \label{eq:level-occu-ratio}
\end{equation}
where $g_1/g_0=3$ is the degeneracy of levels, $T_*=hc/{k_\mathrm{B}\lambda_{21}} = 0.068$K, is the temperature corresponding to the 21cm emission in the atom's rest frame. The spin temperature $T_S$ decides the abundance of singlet and triplet states and couples with different temperatures depending on the underlying physics \cite{Pritchard_2012}. 
At high redshifts ($z<200$) it couples with kinetic temperature $T_K$ of the gas. At redshifts ($z<40$), due to the expansion of the universe, gas temperature density decreases, collision becomes ineffective and $T_S$ couples with the CMB temperature $T_{\gamma}$. After the formation of first luminous objects, $T_S$ couples with the Lyman $\alpha$ photons which correspond to the colour temperature $T_{\alpha}$  \cite{21cm_review,Pritchard_2012, Bharadwaj:2004nr,Ankita_Bera:2022vhw}.

The observable quantity is the excess brightness temperature relative to the CMB and can be expressed as
\begin{equation}
\delta T_\mathrm{b}= \frac{T_S- T_\gamma}{1+z}\tau
\end{equation}
where $\tau\ll 1$ is the reionization depth. After plugging $\tau$ value, we finally arrive at \cite{Pritchard_2012}
\begin{equation}
\delta T_\mathrm{b} \approx 27\left(\frac{\Omega_{\mathrm{b},0}h^2}{0.023}\right)\left(\frac{0.15}{\Omega_{\mathrm{m},0}h^2}\frac{1+z}{10}\right)^{1/2}{x}_\mathrm{HI}(1+\delta_\mathrm{b})\left(\frac{T_S-T_{\gamma}}{T_S}\right)\left[\frac{\partial_rv_r}{(1+z)H(z)}\right]\ \mathrm{mK}\label{eq:Temp_Fluc_21cm}
\end{equation} 
In this equation, $\Omega_{\mathrm{b},0}$ and $\Omega_{\mathrm{m},0}$ are the present values of baryon and total matter density parameters. Neutral fraction of hydrogen is written as $x_\mathrm{HI}$, $\delta_\mathrm{b}$ stands for fractional overdensity in baryons. The last term in the square brackets is the correction due to peculiar velocities along the line of sight (LoS). 


\subsection{Surveys of HI Intensity Mapping \label{sec:HISurvey_IntenMapp}}
To probe the 21cm signal, we employ radio telescopes in two different modes of operation: 
\begin{enumerate}
\item Single Dish Mode (SD): In this mode of operation, we simply add auto-correlation signals from single dishes. After this, we take the Fourier transform. This mode allows us to probe relatively small $k$ values. SKA-Mid is optimized to operate in this mode.
\item Interferometer Mode (IF): In this case, cross-correlated signals from the array elements are combined. This allows for high resolution on small angular scales. We directly get the Fourier transform of the sky. HIRAX and PUMA are designed to operate in this mode.
\end{enumerate}
For line intensity mapping surveys, the thermal noise from the instrument dominates the noise component on the scales of interest, whereas, the shot noise contribution remains minuscule \cite{Gong:2011qf}. So we neglect the short noise contribution for the purpose of this paper  \cite{Karagiannis:2024noise, Santos:2015noise, Bull:2014rha}. 
For surveys in the SD mode, the noise power spectrum of the instrument is \cite{Santos:2015gra}
\begin{equation}
    P_\mathrm{N}^\mathrm{SD}(k_\perp,z) = T^2_\mathrm{sys}(z)\, \chi^2(z)\, \lambda(z)\, \frac{1+z}{H(z)}\, \frac{4\pi f_\mathrm{sky}}{\eta\, N_\mathrm{pol}\, N_\mathrm{dish}\, t_\mathrm{survey}\, \beta^2_\perp(k_\perp, z)} \label{eq:Noise_Power_SD}
\end{equation}
Here the system temperature $T_\mathrm{sys}$ is taken from \cite{SKA:2018ckk}, $\chi$ and $H(z)$ are respectively the comoving distance and Hubble parameter. For SKA-Mid, which is the only SD mode survey considered in this paper, we take efficiency $\eta=1$ and the number of polarizations per feed $N_\mathrm{pol}=2$ following \cite{CosmicVisions21cm:2018rfq}. Further, $\lambda(z)= \lambda_{21}(1+z)$ is the redshifted wavelength of 21cm line and $\lambda_{21}=21$cm is the wavelength in the rest frame of the emitter atom. Also, $\beta_\perp(k_\perp,z)$ is the transverse effective beam which in the Fourier space is \cite{Bull:2014rha}
\begin{equation}
\beta_\perp(k_\perp,z) = \exp\left(-\frac{k_\perp^2 \chi^2(z) \theta_\mathrm{b}^2(z)}{16\ln 2}\right)
\end{equation}
with $\theta_\mathrm{b}(z)=1.22\lambda(z)/D_\mathrm{dish}$ being the FWHM of individual dish beam.
It turns out that the effective beam in the radial direction may be neglected on account of very high frequency resolution of IM experiments \cite{Bull:2014rha}. 

On the other hand, 
the noise power spectrum for interferometer mode is given by \cite{CosmicVisions21cm:2018rfq}
\begin{equation}
    P_\mathrm{N}^\mathrm{IF}(k_\perp,z) = T_\mathrm{sys}^2(z)\,\chi^2(z)\,\lambda(z)\,\frac{(1+z)}{H(z)}\left(\frac{\lambda^2(z)}{A_\mathrm{e}}\right)^2 \frac{2\pi f_\mathrm{sky}}{t_\mathrm{survey}\, n_\mathrm{b}(u,z)\,\theta_\mathrm{b}^2} \label{eq:Noise_Power_IF}
\end{equation}
here $T_\mathrm{sys}$ for PUMA is  taken from \cite{CosmicVisions21cm:2018rfq} and for HIRAX it is the sum of receiver temperature $T_\mathrm{rx}=50$K and sky temperature is taken to be the same as PUMA, the effective area $A_\mathrm{e} = \eta\pi(D_\mathrm{dish}/2)^2$ depends on the efficiency $\eta$. For both HIRAX and PUMA, we take $\eta=0.7$ and $n_\mathrm{b}(u)$ is the baseline number density in the uv plane. Its expression for both HIRAX and PUMA is given in 
the appendix D of \cite{CosmicVisions21cm:2018rfq}. The total survey area is calculated for PUMA using Table \ref{table:Radio_arry}.

\begin{table}
    \centering
    \begin{tabular}{l c c c}
    \hline\hline
     & HIRAX-1024 & PUMA & SKA-Mid \\
        \hline
        Redshift range & $0.775 - 2.55$ & $0.3 - 6.0$ & $0.35 - 3.05$\\
        Integration time, $t_\mathrm{survey}$ (sec) & $1.58\times 10^8$ & $1.58\times 10^8$ & $3.6\times 10^7$ \\
        Sky fraction, $f_\mathrm{sky}$ & $0.36$ & $0.5$  & $0.49$ \\
        Dish diameter, $D_\mathrm{dish}$ (m) & 6 & 6 & 15 \\
        Maximum baseline & 0.25 km & 1.0 km & 150 km \\
        $N_\mathrm{dish}$ & 1,024 & 32,000 & 197 \\
        \hline\hline
    \end{tabular}
    \caption{The instrumental details of the HIRAX-1024 and PUMA radio interferometers used in this work \cite{CosmicVisions21cm:2018rfq,Karagiannis:2022ylq}. Information about SKA-Mid (Band 1) is based on data from the official website \href{https://www.skao.int/en/explore/telescopes/ska-mid}{SKA-MID}.}
    \label{table:Radio_arry}
\end{table}

\subsection{Foreground Problem and its Avoidance \label{sec:forProb_avoid}}
Although, 21cm intensity mapping surveys are extremely advantageous, foreground contamination poses a great challenge. The foreground emission due to galaxies and astrophysical sources turns out to be orders of magnitude larger than the 21cm signal \cite{Alonso:2014dhk, Shaw:2014khi, Spinelli:2021emp}. The foreground emission affects the long wavelength radial Fourier modes, rendering $k_\| < k_{\mathrm{fg}}$ modes inaccessible \cite{Shaw:2013wza, Shaw:2014khi, 21cm_review, Liu_2011A, Liu_2011B}. This is applicable for both single dish and interferometer modes. 
As it turns out that foreground cleaning entails a substantial amount of numerical simulations, thus for our numerical calculations, we use a foreground avoidance approach that imposes a hard cutoff on $k_\|$. In our analysis, we only consider modes having,
\begin{equation}
    |k_\|| \ge k_{\mathrm{fg}}, \quad k_\mathrm{fg}= 0.01h\mathrm{Mpc^{-1}} \label{eq:FGCond1}
\end{equation}
To determine the effect of foreground cut, we compare $k_\mathrm{fg}=0.01h\mathrm{Mpc}^{-1}$ with a bit less optimistic case $k_\mathrm{fg}=0.02h\mathrm{Mpc}^{-1}$. 

In case of interferometer, an additional complication arises on account of foreground leakage to transverse modes. This happens due to chromatic response of the interferometer  \cite{Parsons:2012qh, Pober:2013jna, Seo:2015aza, Pober:2014lva}. It should be emphasized that this isn't a fundamental barrier, rather a technical issue and in principle can be removed with an excellent baseline-to-baseline calibration \cite{Seo:2015aza}. In the present article, however, we incorporate this effect by excluding all modes lying in the foreground wedge. In other words, we impose
\begin{equation}
|k_{\parallel}| > \tan\alpha(z) k_{\perp},\  \label{eq:FGCond2}
\end{equation}
where we have defined
\begin{equation}
\tan(\alpha(z)) = \frac{\chi(z)H(z)\sin[0.61N_W\theta_\mathrm{b}(z)]}{(1+z)}
\end{equation}
$N_W$ represents the primary beam sizes which we take $1$ in this article \cite{Karagiannis:2022ylq}. Writing wedge condition in the form \eqref{eq:FGCond2} allows us to have a nice geometrical interpretation of the foreground conditions 
\cite{Kothari:2023keh} in terms of the Fourier vector components $\mathbf{k}=(k_x,k_y,k_z)$. The condition \eqref{eq:FGCond2} represents a conical region with $k_z$ as symmetry axis. The apex angle of the cone is $\pi/2-\alpha$. Together with \eqref{eq:FGCond1}, the cone becomes truncated. Further, as we increase $z$, $\alpha$ increases, the cone becomes narrower. Thus, the region that we need to exclude on account of foregrounds also becomes smaller. 

In addition to the restrictions imposed by foreground avoidance on the allowed $\mathbf{k}$ values, two more conditions decide the accessible region of the Fourier space.
\begin{enumerate}
\item The first one needed is because we are restricting our analysis in the standard perturbation theory regime, thereby imposing $(k_\|^2+ k_\perp^2)^{1/2} = k \le k_\mathrm{max}$. Above $k_\mathrm{max}$, the theoretical model is considered unreliable \cite{Long:2022dil,Karagiannis:2022ylq}. This selection of $k_\mathrm{max}$ limits the analysis to the reign   
where the description at the tree level matches the numerical results. 
We consider the following non-linear cut off \cite{Volume:2024tvi}
\begin{equation}
k_\mathrm{max}(z) = 0.1(1+z)^{2/(2+n_\mathrm{s})} h\  \mathrm{Mpc}^{-1} \label{eq:NonLin_CutOff}
\end{equation}
where $n_s=0.965$ is the spectral index of the primordial power spectrum. This restriction is also justified by the fact that most of the contribution to the Fisher matrix comes from higher $z$ where the beyond standard perturbation theory effects  are relatively smaller.
\item The other one is $k_\mathrm{min}$, entailing a lower bound on $k$ so that $k\ge k_\mathrm{min}$. This is due to the finite size of the survey volume. It also depends upon the redshift bin width $\Delta z$. 
\end{enumerate}
Imposing all these conditions together renders the Fourier space to take the shape as shown in Figure \ref{fig:FGConds} for single dish (left) and interferometer (right) modes. Also, for the cases we consider, $k_\mathrm{fg}\gg k_\mathrm{min}$, so we don't show $k_\mathrm{min}$ in the figure. Only the blue region will be considered later for our analysis.} 

\begin{figure}
    \centering
\includegraphics[width=7cm]{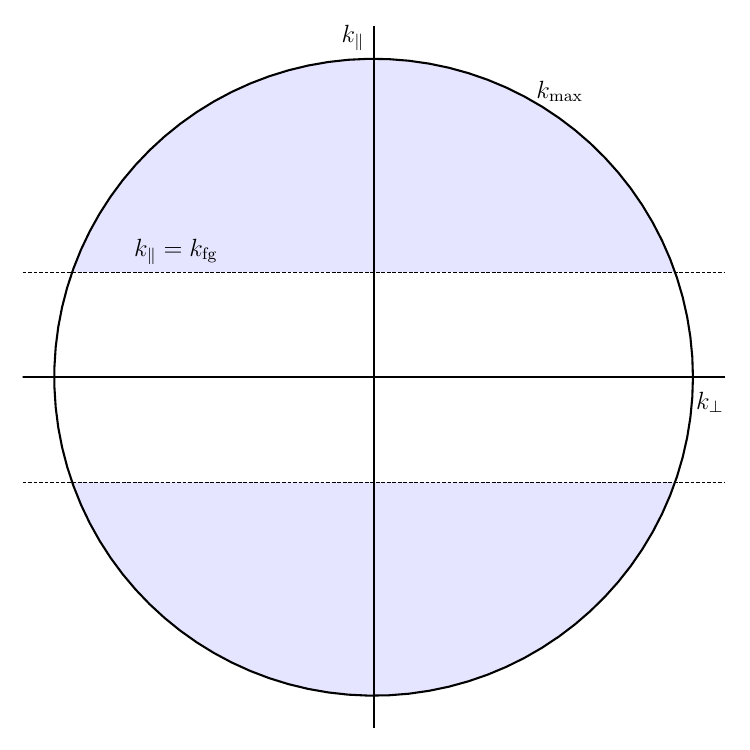}\quad \includegraphics[width=7cm]{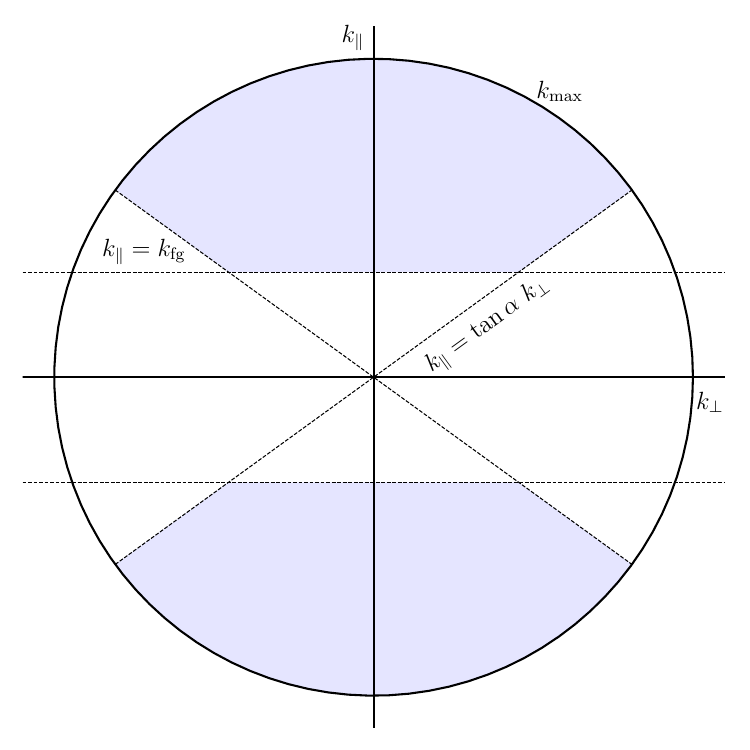}
\caption{Accessible regions (not to scale) of the Fourier space  due to (a) foreground conditions \eqref{eq:FGCond1}, \eqref{eq:FGCond2} (b) non-linear cut-off $k\le k_\mathrm{max}$ \eqref{eq:NonLin_CutOff} represented by the boundary of the larger circle. Since $k_\mathrm{fg}\gg k_\mathrm{min}$, we have not shown the smaller circle in this figure. The \textit{left} panel is for the single dish mode and \textit{right} for the interferometer mode. Only the shaded regions are used in the  Fisher analysis done later in \S\ref{sec:res_method}.}
    \label{fig:FGConds}
\end{figure}


\section{Departure from Rotational Symmetry \label{sec:AsymmTypes}}
In a statistically isotropic and homogeneous Universe, the initial power spectrum of the curvature perturbations from linear perturbation theory can be written as
\begin{equation}
    \langle \zeta(\mathbf{k}) \zeta(\mathbf{k'}) \rangle  = (2\pi)^3 P_\zeta^0(k) \delta_{D}(\mathbf{k+k'})\label{eq:prim_power}
\end{equation}
where $P^0_\zeta(k)$ denotes the isotropic primordial power spectrum of the curvature perturbations in the absence of any asymmetries, Dirac Delta imposes homogeneity and magnitude $|\mathbf{k}|$ dependence of the power spectrum $P^0_\zeta(k)$  represents isotropy. 
However, when rotational symmetries are violated, the primordial power spectrum also becomes a function of the direction $\hat{k}$ and we write $P_\zeta(\mathbf{k})$ in place of $P_\zeta^0(k)$ in \eqref{eq:prim_power}. Nevertheless, the power spectrum still remains homogeneous. In this article, we consider dipolar and quadrupolar asymmetries in the primordial power spectrum.

\subsection{Dipolar Asymmetry \label{sec:Dipolar_Asymmetry}}
Large-scale observations of the CMB show a deviation from the isotropic distribution, demonstrating evidence of anomalies that conflict with the cosmological principle. 
As discussed above, the observations of WMAP and Planck indicate statistical anisotropy aka \textit{hemispherical power asymmetry} in the CMB temperature map \cite{Planck:2018jri, Ma:2011ii, Shiraishi:2016omb}. This is modelled as a dipolar modulation in the statistically isotropic temperature field $T^\mathrm{iso}$ as $T(\hat{\mathbf{n}}) = T^\mathrm{iso}(\hat{\mathbf{n}})(1+A\hat{\mathbf{n}}\cdot\hat{\mathbf{p}})$, where  $\hat{\mathbf{p}}$ is the preferred direction and $\hat{\mathbf{n}}$ is the observation direction. The asymmetry magnitude $A\sim 0.07$ and the direction in galactic coordinates is $\hat{\mathbf{p}}=(l,b)=(205^\circ,-20^\circ)$ \cite{Planck:2019evm}. Phenomenologically, this power asymmetry can be studied by introducing a position dependent dipolar anisotropy in the primordial curvature perturbations {\cite{Shiraishi:galaxy, Shiraishi:2016omb}}. Some papers also explain the effect by introducing inhomogeneity in the primordial power spectrum \cite{Kothari:2015xva, Rath:2014cka}. Given large-scale nature of dipolar anisotropy, it may have connections with inflationary physics. In this regard several, theoretical models have been studied \cite{Byrnes_Adi_Di:2015dub, Lyth:dip_inf, Lyth:14ell}. Some of these models have been ruled out because of their isotropic predictions \cite{Contreras_DiRO:2017, dipolar_rej}. 

A position dependent dipolar anisotropy in the primordial curvature perturbations can be induced by a wavelength longer than the scale of the last-scattering surface in the presence of a large enough primordial bispectrum \cite{Shiraishi:galaxy, Byrnes_Adi_Di:2015dub, Dipolar_Marc} and leads to the the following primordial curvature power spectrum 
\begin{equation}
    P_{\zeta}(k,\hat{\mathbf{n}}) \approx P_{\zeta}^0(k) \Big( 1 + 2A(k) \hat{\mathbf{p}} \cdot \hat{\mathbf{n}} \Big) \label{eq:dipolar_asymm}
\end{equation}
where $\hat{\mathbf{p}}$ is the direction of asymmetry, $\hat{\mathbf{n}}$ the vector along the LoS which we choose to be along Z axis for convenience and $P_{\zeta}^0(k)$ was defined above. Also, we have taken a scale dependent modulation $A(k)$, motivated by the fact that the hemispherical power asymmetry in CMB is present at large angular scales and typically dies out beyond $\ell\gtrsim64$ \cite{Planck:2013lks}. The scale dependence is also suggested by quasar density, where at low $z$ the dipole modulation vanishes at scales $k \sim 1\ \mathrm{Mpc}^{-1}$ \cite{Hirata:2009ar}. Parameterizing the scale dependent amplitude as $A(k) = A^{\prime}F(k)$, with $A'$ denoting the amplitude, we can write \eqref{eq:dipolar_asymm} as
\begin{equation}
    P_{\zeta}(k,\hat{\mathbf{n}}) \approx P_{\zeta}^0(k)[1+ A_0F(k)] \label{eq:Dipolar_Asymm}
\end{equation} 
where $A_0 = 2A^{\prime} \hat{p} \cdot \hat{n} = 2A^{\prime} \hat{p} \cdot \hat{z}$ is the parameter we later constrain using Fisher formalism. 
Further, for the present study, we consider the following scale dependencies $F(k)$ 
\begin{equation}
    F(k) = \left(1-\frac{k}{k_A}\right)^n,\ n=0,2 \label{eq:ansatz_dipolar}
\end{equation}
where $k_A = 1\ \mathrm{Mpc}^{-1}$. The case $n=0$ corresponds to no scale dependence that we compare with $n=2$ used in \cite{Shiraishi:2016omb} to discuss the detectability of 21cm power spectrum. This is also motivated by some theoretical models where varying spectral index $n_\mathrm{s}$ can destroy asymmetry at some scale $k_A$ \cite{Dipolar_Marc}. In addition to this, from CMB observations, a decaying behaviour is captured by the function
\begin{equation}
    F(k) = \left(\frac{k}{k^c_A}\right)^{-0.5}
\end{equation}
where $k_A^c = 0.005\ \mathrm{Mpc}^{-1}$ \cite{Shiraishi:galaxy,Byrnes:2016uqw,Kanno:2013ohv}. We use all three parameterisations for our Fisher analysis results.

\subsection{Quadrupolar Asymmetry \label{sec:quadru_asymmet}}
Although statistical isotropy is the prediction of standard inflationary theory, theoretical research has shown that inflationary models can be constructed that deviate from it. One such model of primordial anisotropy is when the inflaton interacts with a vector field having a nonzero vacuum expectation, transferring the anisotropy \cite{Bartolo:2012sd}. 
Such a model provides a quadrupolar anisotropic primordial power spectrum. We consider the primordial scalar power spectrum to have the following quadrupolar directional dependence about some axis $\hat{\mathbf{d}}$ \cite{Ma:2011ii, Planck:2018jri, Ackerman:2007nb}. In the coordinate system LoS is parallel to $\hat{z}$.


\begin{equation}
P_{\zeta}(\mathbf{k}) \approx  P_{\zeta}^{0}\left(k\right)\left[1+\sum_{M} g_{2 M}(k) Y_{2M}(\hat{k})\right],\  g_{2M}(k)=\frac{8\pi}{15}g_{*}g(k)Y_{2M}^*(\mathbf{\hat{d}}) \label{eq:Quadru_Asymm_Unsimp}
\end{equation}
Then owing to the fact that the power spectrum is real, this expression can be written as
\begin{equation}
    P_\zeta(\mathbf{k}) \approx P^0_\zeta \left[1+ g(k)\sum_i g_i \mathfrak{F}_i(\mu,\phi)\right] \label{eq:Quadru_Asymm}
\end{equation}
real parameters\footnote{Of course, there is no relation between $g_i$'s in \eqref{eq:g-def} \& $g_0$ and $g_1$ used in \eqref{eq:level-occu-ratio} that represented hyperfine level degeneracies used before.}  $g_i$ are chosen for ease of calculations, $\mu=\mathbf{\hat{k}}\cdot \hat{z}$ and $\phi$ is the polar angle of $\mathbf{k}$
\begin{align}
    &g_1 = \sqrt{\frac{8\pi}{15}} g_* \Re(Y_{22}^*(\mathbf{\hat{d}})),\ g_2 = -\sqrt{\frac{8\pi}{15}} g_*\Im(Y_{22}^*(\mathbf{\hat{d}})),\ {g_3 = -\sqrt{\frac{32\pi}{15}} g_* \Re(Y_{21}^*(\mathbf{\hat{d}}))},\notag\\& {g_4 = \sqrt{\frac{32\pi}{15}} g_* \Im(Y_{21}^*(\mathbf{\hat{d}}))},\ g_5 = \sqrt{\frac{4\pi}{45}} g_* Y_{20}(\mathbf{\hat{d}}) \label{eq:g-def}
\end{align}
here $\Re(z)$ and $\Im(z)$ respectively denote the real and imaginary parts of a complex number $z$ and functions $\mathfrak{F}_i$ are defined as
\begin{align}
    &\mathfrak{F}_1 = (1-\mu^2)\cos2\phi,\ \mathfrak{F}_2 = (1-\mu^2)\sin2\phi, \ \mathfrak{F}_3 = \mu\sqrt{1-\mu^2}\cos\phi,\notag\\ &\mathfrak{F}_4 = \mu\sqrt{1-\mu^2}\sin\phi,\ \mathfrak{F}_5 = 3\mu^2-1 \label{eq:frakF-dependence}
\end{align} 
In the present work, we consider the following scale dependencies $g(k)$ \cite{Planck:2018jri}
\begin{equation}
    g(k)=\left(\frac{k}{k_*}\right)^n,\ n=0,\pm1,\pm2 \label{eq:ansatz_quadru}
\end{equation}
here $k_*=0.05 \mathrm{Mpc}^{-1}$ is the pivot scale chosen by Planck \cite{Planck:2018jri}. 
We later constrain parameters $g_i$ using Fisher formalism.

\section{Spectra in the presence of Asymmetries \label{sec:Power_Bispec}}
Cosmological constraints are usually given with the help of power spectrum which is the Fourier transform of the 2 point correlation function (PCF). When the primordial curvature perturbation is considered as Gaussian random field and its evolution is assumed to be linear, the 2PCF is sufficient to extract all the relevant information. In such a case, it is known that all odd correlations are zero and all even correlators are expressible in terms of the 2PCF. However, in the presence of non-gaussianities that can be either of (a) the primordial type or (b) arising due to non-linear structure formation on account of gravitational instabilities \cite{Scoccimarro:1999ed,Fry1984}, higher order correlators might provide more information. The next higher order statistic is bispectrum which is the Fourier transform of the 3PCF, and is a measure of non-gaussianities. This statistic has been shown to be useful in improving parameter constraining power and breaking degeneracies \cite{Cosmo_bispectra:2006pa}. In our work, we consider the initial perturbations $\zeta$ to be gaussian and study non-gaussianities due to non-linearities. Next, we present the expressions for both power and bispectrum in the presence of primordial anisotropies discussed in \S \ref{sec:AsymmTypes}.

\subsection{Power Spectrum}

The 21cm power spectrum $P_{21}(\mathbf{k},z)$ is defined through 
2PCF of Fourier coefficient $\delta T_\mathrm{b}(\mathbf{k}, 
z)$ of the excess brightness temperature $\delta T_\mathrm{b}(\hat{\mathbf{n}},z)$, (c.f. \eqref{eq:Temp_Fluc_21cm})
\begin{equation}
\langle \delta T_\mathrm{b}(\mathbf{k},z)\delta T_\mathrm{b} (\mathbf{k^{\prime}},z)\rangle = (2\pi)^3P_{21}(\mathbf{k},z)\delta_{D}(\mathbf{k}+\mathbf{k^{\prime}}) \label{eq:21cm_Pow_Spec}
\end{equation} 
assuming the LoS is along $\hat{z}$ axis and primordial anisotropies are absent, the expression for 21cm differential brightness temperature power spectrum $P_{21}$ is 
\begin{equation}
P_{21}^0(\mathbf{k},z)\equiv P_{21}^0(k,\mu,z)=\Bar{T}^2_\mathrm{b}(z)D_\mathrm{FoG}(\mathbf{k},z)(b_\mathrm{1}(z)+\mu^2f)^2P_\mathrm{m}^0(k,z) + P_\mathrm{N}(\mathbf{k},z) \label{eq:power_sp_eq}
\end{equation}
where $\bar{T}_\mathrm{b}$ is the mean brightness temperature, the term $D_\mathrm{FoG}(\mathbf{k}) = \text{exp}[-k^2\mu^2\sigma^2_\mathrm{P}(z)]$ corresponds to the \textit{Fingers of God} (FoG) effect, with $\sigma_\mathrm{P}$ denoting the LoS dispersion in relative galaxy velocities. In this paper, we use $\sigma_\mathrm{p}=350/H(z)$, as per Ref.
\cite{FoG:1996cd}. Further,  $b_1$ is the neutral hydrogen bias given by \cite{Jolicoeur:2020eup}
\begin{equation}
    b_1(z) = 0.754 + 0.0877z + 0.0607z^2-0.00274z^3 \label{eq:b1_bias}
\end{equation}
and $\mu= \hat{k}\cdot \hat{z}$ is the cosine of angle {between $\hat{k}$ and} LoS, $f$ is the linear growth rate related to growth factor $D$ by $f=d\ln{D(a)}/d\ln{a}$. The quantity, $P_\mathrm{m}^0({k},z)$ denotes the linear matter power spectrum in the absence of any anisotropies. 

As was discussed before, the noise power spectrum $P_\mathrm{N}(\mathbf{k},z)$ doesn't get any contribution from the shot noise.   The only contribution comes from the instrumental noise $P_\mathrm{N}(\mathbf{k},z)$ with the expressions given in \eqref{eq:Noise_Power_SD} and \eqref{eq:Noise_Power_IF} for the single dish and interferometer modes of operation. 
 
The matter power spectrum $P_\mathrm{m}^0$ appearing in \eqref{eq:power_sp_eq} is related to the power spectrum of the primordial curvature perturbations $P_\zeta^0$ as \cite{CosmicVisions21cm:2018rfq} 
\begin{equation}
    P_\mathrm{m}^0(k,z) = \mathcal{T}^2(k,z)P_{\zeta}^0(k)
\end{equation}
where $\mathcal{T}(k)$ is the relevant transfer function and $0$ in the superscripts are the quantities in the absence of primordial asymmetries. In the presence of asymmetries, this equation gets modified to
\begin{equation}
    P_\mathrm{m}^0(k,z) \longrightarrow P_\mathrm{m}(\mathbf{k},z) = \mathcal{T}^2(k,z)P_{\zeta}(\mathbf{k}) \label{eq:Symm2Asymm_MattPow}
\end{equation}
where we assume that only the primordial curvature power spectrum has the direction dependence and the transfer functions are still direction-independent. Using expressions of dipolar \eqref{eq:Dipolar_Asymm} and quadrupolar \eqref{eq:Quadru_Asymm} asymmetries in \eqref{eq:Symm2Asymm_MattPow}, we conclude that the effect of introducing asymmetry is obtained by the following substitutions in the isotropic matter power spectrum 
\begin{align}
    \mathrm{Dipolar} &: P_\mathrm{m}^0(k,z) \longrightarrow P_\mathrm{m}(\mathbf{k},z)=  P_\mathrm{m}^0(k,z)  [1 + A_0F(k)] \label{eq:Dipolar_Substitute} \\
    \mathrm{Quadrupolar} &: P_\mathrm{m}^0(k,z) \longrightarrow 
 P_\mathrm{m}(\mathbf{k},z)= P_\mathrm{m}^0(k,z)\Big[1 +g(k)\sum_i g_i \mathfrak{F}_i(\mu,\phi)\Big]\label{eq:Quadrupolar_Substitute}
\end{align}
where the scales dependencies $F(k)$ and $g(k)$ are given in \eqref{eq:ansatz_dipolar} and \eqref{eq:ansatz_quadru} and $\mathfrak{F}$ are given in \eqref{eq:frakF-dependence}. 
{Recently, it has been pointed out in \cite{Shiraishi:2023zda, 2024arXiv240912004M} that due to non-zero 
quadrupolar anisotropy, 
expression  \eqref{eq:Quadrupolar_Substitute} should also contain a 
bias term, say $b_1^{(2)}$, corresponding to a tidal field at linear order.} 
{Ref \cite{2024arXiv240912004M} discusses $b_\mathrm{h}^{(2)}$ in this context while relating halo overdensity with dark matter density fluctuations. The authors estimate this parameter using simulations and find that the bias factor $b_\mathrm{h}^{(2)}$ is nearly scale-invariant on linear scale. We believe such a term is important for a more careful analysis but would require an estimation of the bias factor $b_1^{(2)}$. Due to present unavailability of such 
results, 
we neglect the contribution arising from this term for both power spectrum and bispectrum analysis.}

Thus, our final expressions for the 21cm power spectrum in the presence of the dipolar and quadrupolar anisotropies are simplified to
\begin{align}
P_{21}^\mathrm{D}(\mathbf{k},z,\hat{\mathbf{n}})&= P^0_{21}(\mathbf{k},z)+ A_0(P^0_{21}(\mathbf{k},z) - P_\mathrm{N}(\mathbf{k},z))F(k)  \label{eq:Final_Power_Dipole} \\
P_{21}^\mathrm{Q}(\mathbf{k},z)&=P^0_{21}(\mathbf{k},z)+ g(k)\sum_i g_i \mathfrak{F}_i(\mu,\phi)(P^0_{21}(\mathbf{k},z) - P_\mathrm{N}(\mathbf{k},z))  \label{eq:Final_Power_Quadru}
\end{align}
\subsection{Bispectrum}
In terms of the Fourier coefficient $\delta T_\mathrm{b}(\mathbf{k}, 
z)$ of the excess brightness temperature  $\delta T_\mathrm{b}(\hat{\mathbf{n}},z)$, i.e., equation \eqref{eq:Temp_Fluc_21cm}, the 21cm bispectrum $B_{21}$ is defined as 
\begin{equation}
\langle\delta T_\mathrm{b}(\mathbf{k}_1,z) \delta T_\mathrm{b}(\mathbf{k}_2,z) \delta T_\mathrm{b}(\mathbf{k}_3,z)\rangle = (2\pi)^3\delta_D(\mathbf{k}_1+ \mathbf{k}_2 + \mathbf{k}_3) B_{21}(\mathbf{k}_1, \mathbf{k}_2, \mathbf{k}_3, z) \label{eq:21Bispec_Def}
\end{equation}
The Dirac Delta imposes (a) the condition that the vectors $\mathbf{k}_1$, $\mathbf{k}_2$ and $\mathbf{k}_3$ form three sides of the triangle and (b) homogeneity as before. Apparently, there are 9 variables (three components each for $\mathbf{k}_i$, $i=1,2,3$) in the bispectrum expression \eqref{eq:21Bispec_Def} but three variables are eliminated by the Dirac Delta condition $\mathbf{k}_1 + \mathbf{k}_2 + \mathbf{k}_3=0$, thereby leading to 6 independent variables. Also, the bispectrum is invariant under a rotation in the plane perpendicular to the LoS \cite{SinghGill:2024oio}. So we are left with 5 variables that completely describe the bispectrum. We can choose three variables to define the shape of the triangle and remaining two for describing its orientation with respect to the LoS. Thus finally, 21cm bispectrum, \eqref{eq:21Bispec_Def} can be written as $B_{21}(\mathbf{k}_1, \mathbf{k}_2, \mathbf{k}_3, z)=B_{21}(k_1,k_2,k_3,\mu,\phi,z)$, where $k_1$, $k_2$ and $k_3$ are the sides of the triangle, $\mu$ is the cosine of angle between $\mathbf{k}_1$ and Z axis (the chosen LoS). Also, $\phi$ is the azimuthal angle about $\mathbf{k}_1$ \cite{Scoccimarro:1999ed} (see discussion in Appendix \ref{sec:azimutha-and-polar}). 

In whatever follows, we take $\mu_i = \hat{\mathbf{k}}_i \cdot \hat{z}$, the cosine of the angle between the vector $\mathbf{k}_i$ and the Z axis. These are expressible in terms of $\mu$ and $\phi$ \cite{Karagiannis:2022ylq}. The tree level bispectrum in the redshift space when primordial asymmetries are absent is
\begin{equation}
B_{21}^0(\mathbf{k}_1,\mathbf{k}_2,\mathbf{k}_3,z) = {\Bar{T}_\mathrm{b}^3(z)}\Big\{D_\mathrm{FoG}^B\Big[2Z_1(\mathbf{k}_1,z)Z_1(\mathbf{k}_2,z)Z_2(\mathbf{k}_1,\mathbf{k}_2,z)P_\mathrm{m}^0 ({k}_1,z)P_\mathrm{m}^0 ({k}_2,z) + 2\text{perm}\Big]\Big\} \label{eq:Bispectrum_main}
\end{equation}
In this equation, the FoG effect for bispectrum is $D_\mathrm{FoG}^{B}(\mathbf{k}_1,\mathbf{k}_2,\mathbf{k}_3,z) = \text{exp}[-\sigma_\mathrm{B}^2(z)\sum_{i=1}^3(k_i\mu_i)^2]$, $Z_i$ is the redshift dependent kernel at $i$th order in perturbation theory {and $\sigma_\mathrm{B} = \sigma_\mathrm{P}$  \cite{Karagiannis:2019jjx}}. Assuming gaussian initial conditions, the expressions for the kernels can be written as
\begin{align}
    Z_1(\mathbf{k}_1,z) &= b_1(z) + f(z)\mu_1^2 \\
    Z_2(\mathbf{k}_1,\mathbf{k}_2,z) &= b_1(z)F_2(\mathbf{k}_1,\mathbf{k}_2) + f(z)\mu_{12}^2G_2(\mathbf{k}_1,\mathbf{k}_2) + \frac{b_2(z)}{2} + \frac{b_{s^2}(z)}{2}S_2(\mathbf{k}_1,\mathbf{k}_2) \notag \\
    &\quad + \frac{1}{2}f(z)\mu_{12}k_{12}\left[\frac{\mu_1}{k_1}Z_1(\mathbf{k}_2,z) + \frac{\mu_2}{k_2}Z_1(\mathbf{k}_1,z) \right]
\end{align}
here $\mu_{12}=(\mu_1k_1+\mu_2k_2)/k_{12}$, $k_{12}^2=(\mathbf{k}_1+\mathbf{k}_2)^2$,  $F_2(\mathbf{k}_1,\mathbf{k}_2)$ \&  $G_2(\mathbf{k}_1,\mathbf{k}_2)$ are the mode-coupling part of the matter density contrast \& matter velocity \cite{LSS_SPT:2001qr} and $S_2(\mathbf{k}_1,\mathbf{k}_2) = (\hat{k}_1\cdot\hat{k}_2)^2-1/3$ is the tidal kernel \cite{Tidal:2012hs}. 
The neutral hydrogen biases $b_i(z)$, $i=1,2$ and $b_{s^2}$ relate the statistics of observed tracers with the dark matter distribution.
The bias $b_1$ was given in \eqref{eq:b1_bias}, $b_2$ is given by the fitted formula \cite{Jolicoeur:2020eup}
\begin{equation}
    b_2(z) = -0.308-0.0724z-0.0534z^2+0.0247z^3
\end{equation}
and the tidal field bias is related to $b_1$ as $b_{s^2} = -4(b_1-1)/7$ \cite{bias_fit:2023rgp,Karagiannis:2022ylq,Tidal_bias:2012hs}. We can now obtain the 21cm bispectrum in the presence of dipolar anisotropies by substituting \eqref{eq:Dipolar_Substitute} in \eqref{eq:Bispectrum_main}
\begin{align}
   B_{21}^\mathrm{D}(\mathbf{k}_1, \mathbf{k}_2, \mathbf{k}_3, \hat{\mathbf{n}}, z) &= \bar{T}_b^3(z) D_\mathrm{FoG}^\mathrm{B} \Big[ 2 Z_1(\mathbf{k}_1, z) Z_1(\mathbf{k}_2, z) Z_2(\mathbf{k}_1, \mathbf{k}_2, z) P_\mathrm{m}^0(k_1, z)P_\mathrm{m}^0(k_2, z) \notag \\
   &\times [1 + A_0 (F(k_1)+F(k_2))] + 2\ \text{perm} \Big] \label{eq:bispectrum_dip}
\end{align}
Similarly, for quadrupolar anisotropy, the following result is obtained using \eqref{eq:Quadrupolar_Substitute} in \eqref{eq:Bispectrum_main}

\begin{align}
B_{21}^\mathrm{Q}(\mathbf{k}_1,\mathbf{k}_2,\mathbf{k}_3,z) &= \bar{T}_\mathrm{b}^3(z)D_\mathrm{FoG}^\mathrm{B} \Big\{ 2 Z_1(\mathbf{k}_1,z) Z_1(\mathbf{k}_2,z) Z_2(\mathbf{k}_1,\mathbf{k}_2,z) P_\mathrm{m}^0(k_1,z) P_\mathrm{m}^0(k_2,z)  \notag \\
&\times \big[1 + \sum_i g_i \{g(k_1)\mathfrak{F}_i(\mu_1,\phi_1)+g(k_2)\mathfrak{F}_i(\mu_2,\phi_2)\}\big] + 2 \text{perm} \Big\} \label{eq:bispectrum_quad}
\end{align} 
In \eqref{eq:bispectrum_dip} and \eqref{eq:bispectrum_quad}, we assume both dipolar and quadrupolar asymmetry parameters are small so that we only  need to consider first-order terms $\mathcal{O}(A_0)$ and $\mathcal{O}(g_i)$. Also, the polar angles $\phi_i$ in \eqref{eq:bispectrum_quad} are calculated in Appendix \ref{sec:azimutha-and-polar}. 



\section{Results and Discussion \label{sec:res_method}}
In this section, we use Fisher formalism \cite{Long:2022dil, Karagiannis:2022ylq, amendola2024} to constrain dipolar ($A_0$) and quadrupolar ($g_i$, $i=1,2,\ldots,5$) asymmetry parameters. As we've only one parameter in case of dipolar asymmetry, the Fisher matrix will have only one element for either power spectrum or bispectrum. For the quadrupolar case, the size of the matrix depends upon whether we evaluate power spectrum or bispectrum. As we discuss later, we find that in case of power spectrum, $g_1$ shows degeneracy with $g_2$ and $g_3$ with $g_4$. Thus effectively there are only 3 parameters to constrain when we consider power spectrum. We find no such simplification in case of bispectrum and need to constrain all 5 parameters. 

\begin{figure}[t]
\centering
\includegraphics[width=0.32\textwidth]{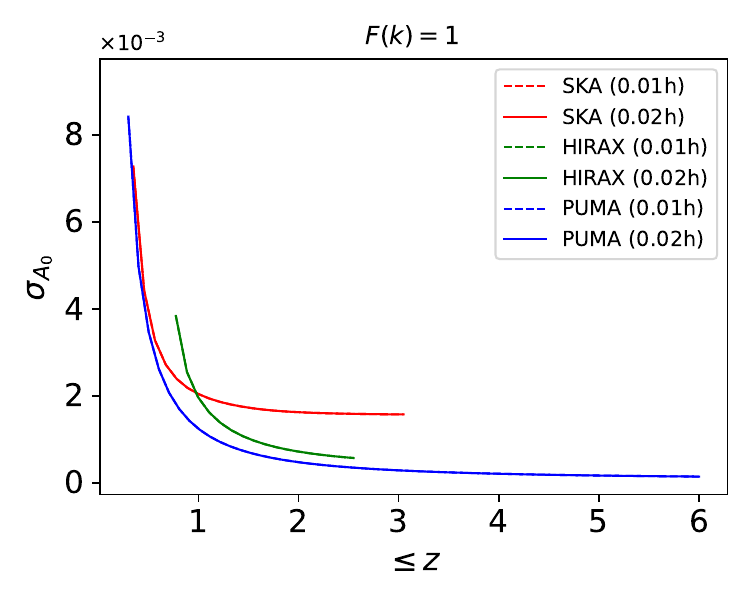}
\includegraphics[width = 0.32\textwidth]{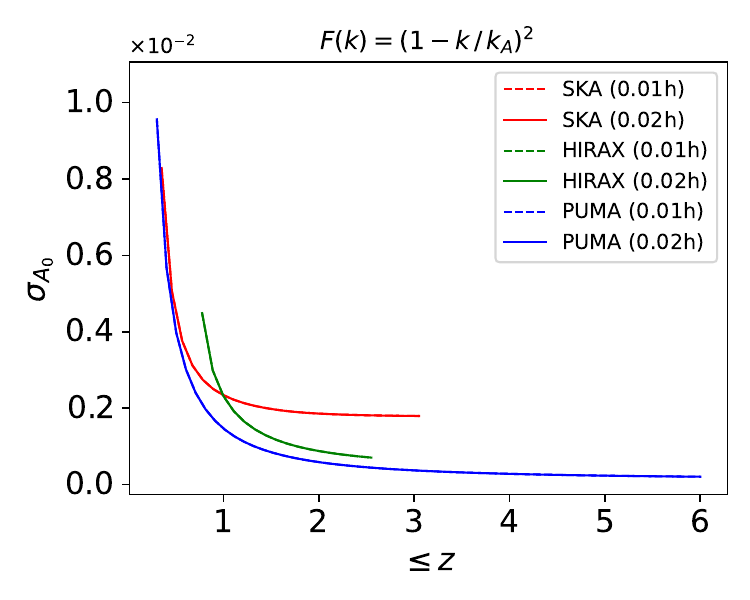}
\includegraphics[width=0.32\textwidth]{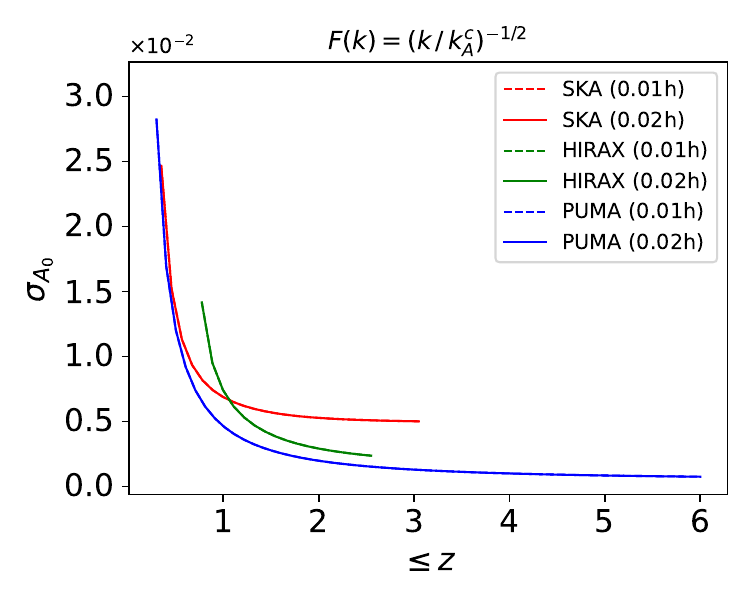}
\caption{Cumulative error $\sigma_{A_{0}}$ on the dipolar anisotropy parameter $A_0$ using 21cm power spectrum for SKA-Mid, HIRAX and PUMA surveys. The results are shown with the redshift range relevant for the considered survey (see Table \ref{table:Radio_arry}). We have shown our results for both $k_\mathrm{fg}= 0.01h\mathrm{Mpc}^{-1}$ and $ 0.02h\mathrm{Mpc}^{-1}$. First two plots correspond to the scaling $F(k)=(1-k/k_A)^n$ for $n=0,2$ and $k_A = 1\ \mathrm{Mpc}^{-1}$ while the rightmost one is for $F(k) = (k/k^c_A)^{-0.5}$ with $k_A^c = 0.005\mathrm{Mpc}^{-1}$ (see \S \ref{sec:Dipolar_Asymmetry} for more details). For a given $z$ in the common range, PUMA gives the best constraints irrespective of the chosen scale dependence.}
\label{fig:Dipolar_Power_z_kmaxFix}
\end{figure}

\begin{figure}[t]
\centering
\includegraphics[width=0.32\textwidth]{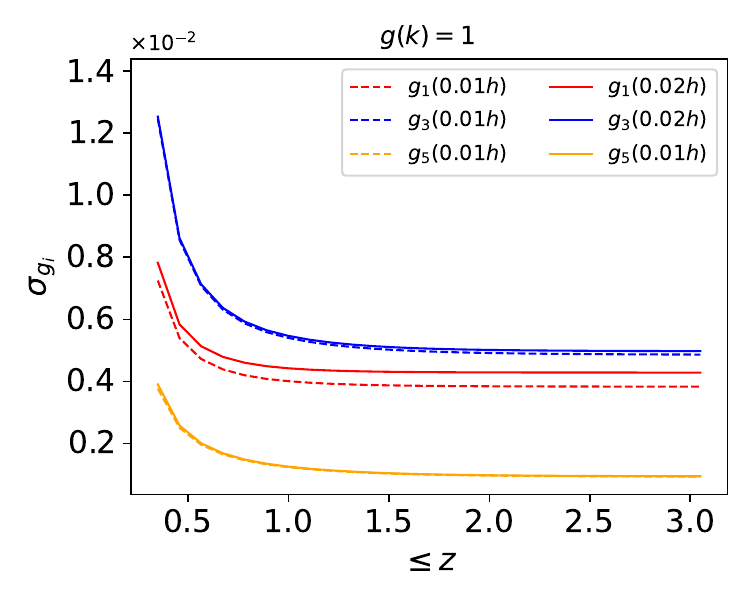}
\includegraphics[width=0.32\textwidth]{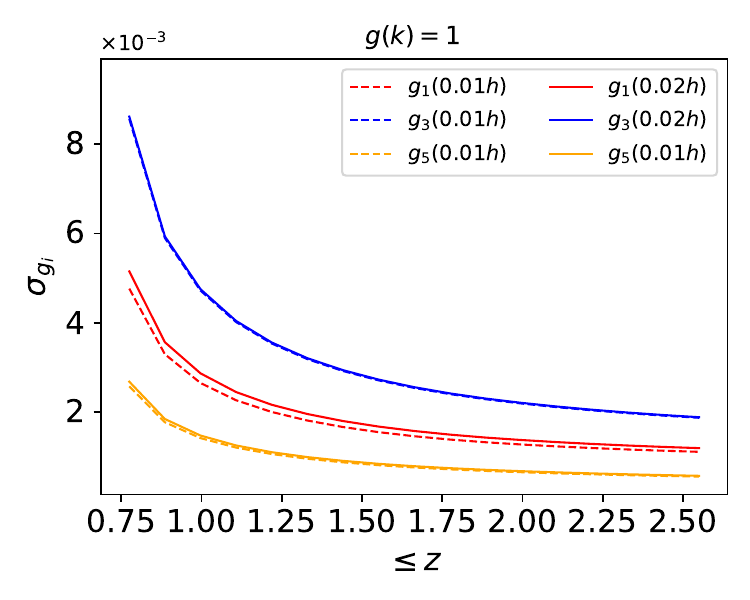}
\includegraphics[width = 0.32\textwidth]{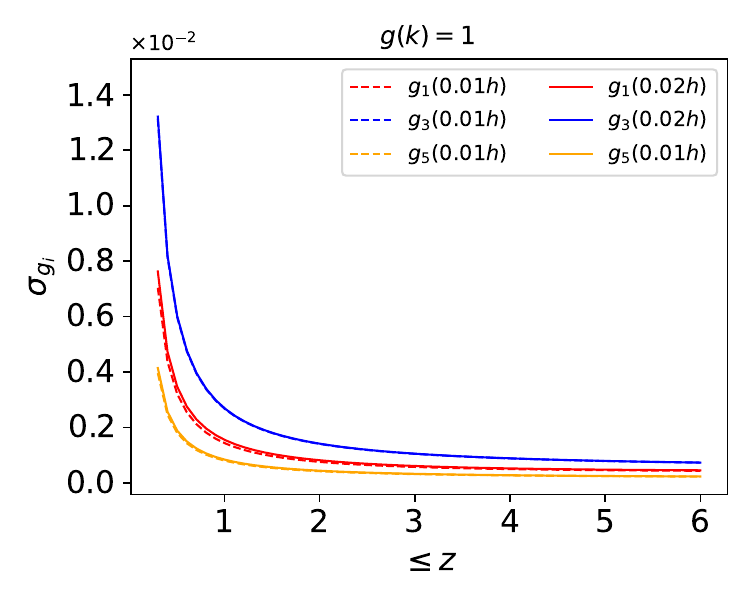}
\caption{Cumulative error $\sigma_{g_{i}}$ on the quadrupolar anisotropy parameters $g_{i}$ using power spectrum for SKA, HIRAX and PUMA surveys (respectively from left to right). The results are shown with the range relevant for the considered survey (see Table \ref{table:Radio_arry}) for both $k_\mathrm{fg}= 0.01h\mathrm{Mpc}^{-1}$ and $0.02h\mathrm{Mpc}^{-1}$. We've chosen the scale dependence $g(k) = 1$ in these plots. As can be seen, the best constraints are obtained on $g_5$.  
}
\label{fig:Quadru_Power_g1_all_surveys}
\end{figure}

\begin{figure}[t]
\centering
\includegraphics[width=0.40\textwidth]{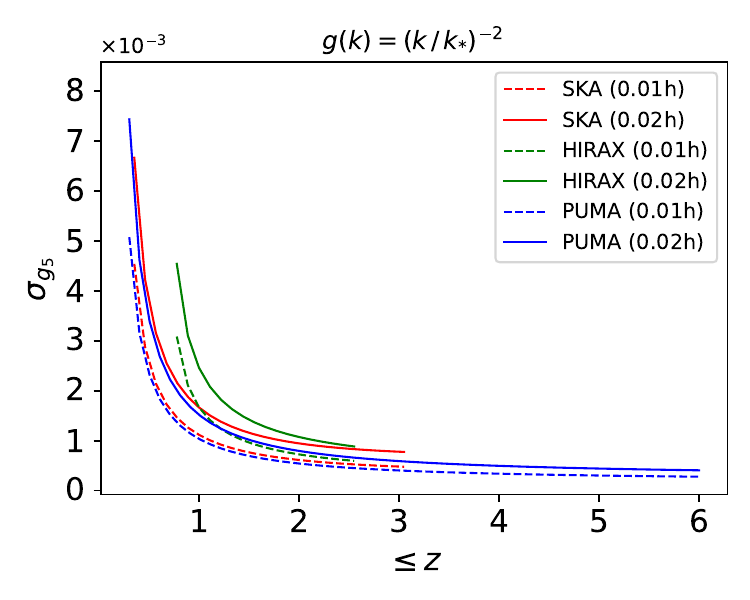}
\includegraphics[width=0.40\textwidth]{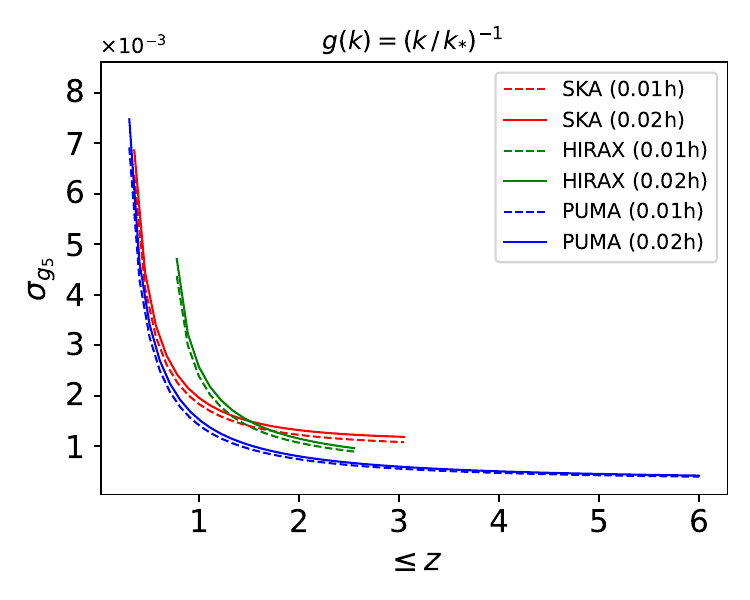} \\
\includegraphics[width=0.4\textwidth]{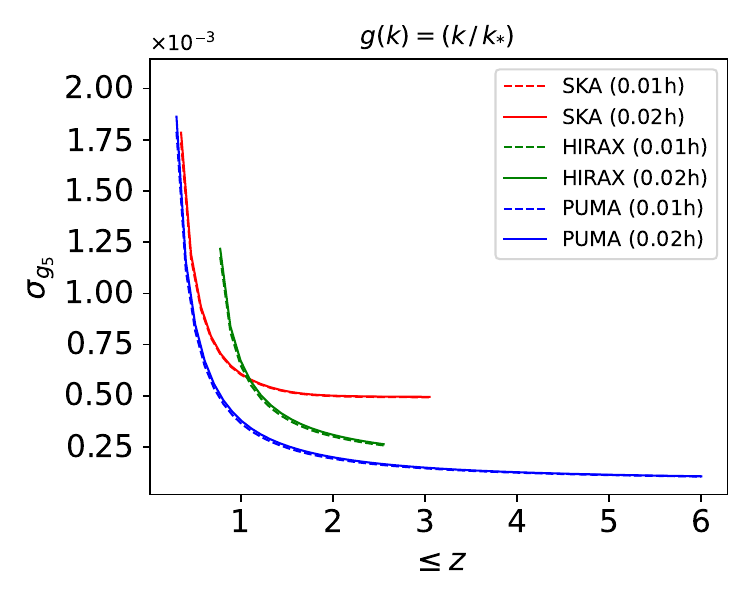}
\includegraphics[width=0.4\textwidth]{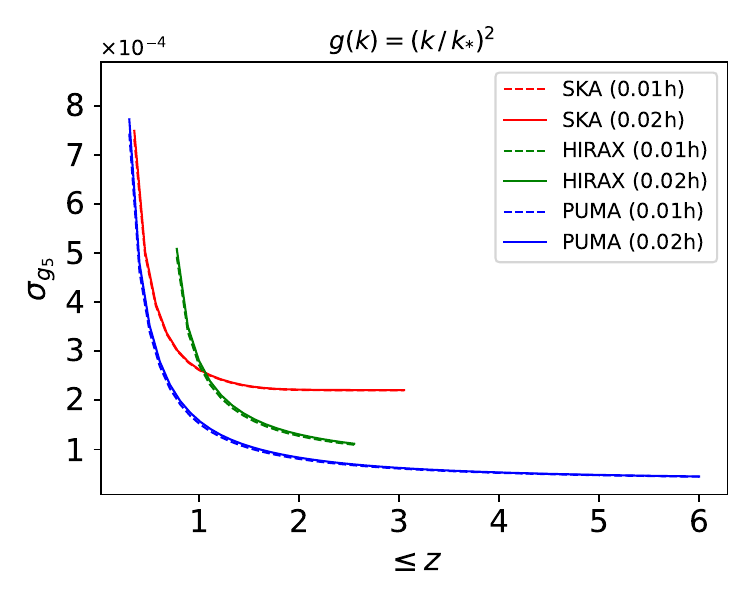}
\caption{Cumulative error $\sigma_{g_5}$ on the quadrupolar anisotropy parameter $g_{5}$ using 21cm power spectrum for SKA-Mid, HIRAX and PUMA. In these plots, we focus exclusively on $g_5$, we find to be the best  constrained parameter. The results are shown for various scalings $g(k)=(k/k_*)^n$, discussed in \S\ref{sec:quadru_asymmet}. \textit{Top panel:} From left to right, $n=-2,-1$ and \textit{bottom panel:} $n=1,2$. Just like for dipolar anisotropy, in this case too, PUMA gives the best constraints in the common range.}
\label{fig:Quadru_powerZ_kmaxFix}
\end{figure}

\subsection{Power Spectra Results}\label{subsec:Power_spectrum_result}
The Fisher matrix element $F^{D}_P(z)$ from power spectrum for dipolar anisotropy is
\begin{equation}\label{eq:Fisher_matrix}
F^{D}_P(z)=\frac{V_\mathrm{sur}}{16\pi^3}\int k^2\ \d k\, \d\mu\, \d \phi\left(\frac{\partial P_{21}^{D}(\mathbf{k},z)}{\partial A_0}\right)^2 \frac{1}{(P^{D}_{21}(\mathbf{k},z))^2}
\end{equation}
where the integration is performed over the restricted region of the Fourier space on account of (a) foregrounds, (b) non-linearities (see \S \ref{sec:HISurvey_IntenMapp}) and (c) survey volume. This region also depends upon the mode of operation -- single dish or interferometer and is depicted in Figure \ref{fig:FGConds} for both modes.
The 21cm power spectrum in the presence of dipolar anisotropies $P^{D}_{21}(\mathbf{k},z)$ is given in \eqref{eq:Final_Power_Dipole} and   $V_\mathrm{sur}$ is the survey volume \cite{Long:2022dil, Volume:2024tvi}
\begin{equation}
V_\mathrm{sur}=\frac{4\pi}{3}f_\mathrm{sky}\left[\chi^3\left(z+\frac{\Delta z}{2}\right)-\chi^3\left(z-\frac{\Delta z}{2}\right)\right].
\end{equation}
Here $\chi(z)$ is the comoving distance to redshift $z$ and $\Delta z$ is the redshift bin width, which 
{for the present study, we choose $\Delta z= 0.1$ \cite{Volume:2024tvi}.} Physically, the survey volume represents the largest length scale and hence the smallest value of $k\ge k_\mathrm{min}=k_\mathrm{f}$, that we can include in our analysis. If we approximate the survey volume with a cube of size $L$, then $L=V_{\mathrm{sur}}^{1/3}$ and the fundamental frequency  of the survey is then $k_\mathrm{f}=2\pi/L$. Using \eqref{eq:Final_Power_Dipole}, \eqref{eq:Fisher_matrix} simplifies to
\begin{equation}
    F^{D}_P(z) = \frac{V_\mathrm{sur}}{16\pi^3}\int \d k\d\mu \left[\frac{(P_{21}^0(k,\mu,z) - P_N(k,\mu,z))\, k\, F(k)}{P_{21}^0(k,\mu,z)}\right]^2
\end{equation}
Here we have performed integration over $\phi$ variable and also in the denominator we've approximated $P_{21}^D(k,\mu,z) \approx P_{21}^0(k,\mu,z)$, as the dipolar anisotropy is considered to be small compared to $1$. For the quadrupolar asymmetry, we've 5 parameters to constrain, thus the Fisher matrix element $F^Q_{P,ij}$ is written as
\begin{equation}
    F^Q_{P,ij} = \frac{V_\mathrm{sur}}{16\pi^3} \int k^2 \mathrm{d}k \d\mu \d\phi \frac{\partial P_{21}^Q(\mathbf{k},z)}{\partial g_i} \frac{\partial P_{21}^Q(\mathbf{k},z)}{\partial g_j} \frac{1}{(P^Q_{21}(\mathbf{k},z))^2}
\end{equation}
Using \eqref{eq:Final_Power_Quadru} in this equation, we get
\begin{equation}
    F^Q_{P,ij} = \frac{V_\mathrm{sur}}{16\pi^3} \int \d k\d \mu \d \phi\ \mathfrak{F}_i \mathfrak{F}_j \left[\frac{(P^0_{21}(k,\mu,z)-P_\mathrm{N}(k,\mu,z))\, k\, g(k)}{P_{21}^0(k,\mu,z)}\right]^2\label{eq: Quad_Power_fisher}
\end{equation}
The $\phi$ dependence in this integral comes only from the $\mathfrak{F}_i$, see \eqref{eq:frakF-dependence}. An integration over $\phi$ renders all non-diagonal terms zero. 
{However, due to the identical functional dependence $\mu$ in the pair $\mathfrak{F}_1$ and $\mathfrak{F}_2$ \& $\mathfrak{F}_3$ and $\mathfrak{F}_4$, the corresponding Fisher integrals are equal. As a result, the parameters $g_1$ and $g_2$ (and similarly $g_3$ and $g_4$) receive identical constraints. 
Thus for the quadrupolar study using power spectrum, we report constraints only on three 
parameters: $g_1$, $g_3$ and $g_5$.}

We show the power spectrum constraints on dipolar anisotropy parameter in Figure \ref{fig:Dipolar_Power_z_kmaxFix} and those on quadrupolar anisotropy in Figures \ref{fig:Quadru_Power_g1_all_surveys} and \ref{fig:Quadru_powerZ_kmaxFix}. In all these plots, cumulative errors as a function of $z$, considering foreground cuts $k_\mathrm{fg}=0.01h \mathrm{Mpc}^{-1}$ and $=0.02h \mathrm{Mpc}^{-1}$ (see \eqref{eq:FGCond1}) have been considered. As we discussed before, in case of quadrupolar anisotropy, we've only 3 parameters ($g_1$, $g_3$ and $g_5$) to constrain. To avoid clutter, we first check the behaviour of $\sigma_{g_i}$ in three separate plots for all three surveys with $g(k)=1$ in Figure \ref{fig:Quadru_Power_g1_all_surveys}. We find that constraints on $g_1$, $g_3$ and $g_5$ are of the same order $\lesssim\mathcal{O}(10^{-3})$, but the best constraints are found on the parameter $g_5\propto g_{20}$ (see \eqref{eq:Quadru_Asymm_Unsimp} for $g_{2M}$ definition). We find the same trend in case of other scalings $g(k)$, again we've not included those plots to avoid clutter. Our best constrained parameter out of $g_1$, $g_3$ and $g_5$ is $g_5$, thus in Figure \ref{fig:Quadru_powerZ_kmaxFix}, we show only $\sigma_{g_5}$. The plots exhibit the following features.
\begin{enumerate}
    \item The cumulative error $\sigma$ for both dipolar ($A_0$) and best constrained parameter ($g_5\propto g_{20}$) of quadrupolar anisotropies at $z=1$ is $\sim 10^{-3}$ for all three surveys considered.
    \item As expected, power spectrum with a smaller value of $k_\mathrm{fg}=0.01h \mathrm{Mpc}^{-1}$ has a better constraining power as compared to that with $k_\mathrm{fg}=0.02h \mathrm{Mpc}^{-1}$. This is because larger foreground cuts render a larger number of $k$ modes unusable.
    \item Regardless of the parameter ($A_0$ or $g_5$) constrained or the scaling ($F(k)$ or $g(k)$) used, PUMA gives the best constraints in the common redshift range $0.78\le z\le 2.55$.
    \item For the dipolar case, Figure \ref{fig:Dipolar_Power_z_kmaxFix}, we find that the error doesn't change significantly with the scaling. Further, the difference between errors, for the chosen values of $k_\mathrm{fg}$ ($0.01h$ or $0.02h$) isn't very significant. For a given redshift and scaling, the error slightly increases as $n$ increases.
    \item But we see a slightly different behaviour in case of quadrupolar anisotropy parameter $g_5$ in Figures \ref{fig:Quadru_Power_g1_all_surveys} and \ref{fig:Quadru_powerZ_kmaxFix}. We first notice that the difference between constraints obtained with and without foregrounds keep decreasing as $n$ is increased on $-2$ to $2$. Apart from this, there isn't any specific pattern in error obtained for a given survey and redshift by changing $n$.
\end{enumerate}

\subsection{Bispectra Results}
For the dipolar anisotropy, the Fisher matrix element from bispectrum becomes 
\begin{equation}
F_{B}^{D}(z) = \frac{1}{4\pi}\sum_{k_1k_2k_3}\int\d\mu_1\int_0^{2\pi}\d\phi \left(\frac{\partial B_{21}^D(\mathbf{k}_1,\mathbf{k}_2,\mathbf{k}_3,z)}{\partial A_0}\right)^2\frac{1}{\Delta B^2(\mathbf{k}_1,\mathbf{k}_2,\mathbf{k}_3,z)}
\end{equation}
where $\mu_1$ and $k_i$ ranges are determined as per the conditions discussed before in \S \ref{sec:forProb_avoid}. Additionally, $k_1,k_2,k_3$ are chosen to satisfy (i) the triangle inequality and (ii) $k_\mathrm{min}\le k_3 \le k_2 \le k_1 \le k_\mathrm{max}$. As the departure from isotropy is expected to be small \cite{Cosmo_bispectra:2006pa}, the variance $\Delta B^2$, assuming Gaussian statistics  
and neglecting the non diagonal part is  
\begin{equation}
\Delta B^2(\mathbf{k}_1,\mathbf{k}_2,\mathbf{k}_3,z) = s_{123}\ \pi\ k_\mathrm{f}^3(z)\frac{P_{21}^\mathrm{D}(\mathbf{k}_1,z)P_{21}^\mathrm{D}(\mathbf{k}_2,z) P_{21}^\mathrm{D}(\mathbf{k}_3,z)}{k_1k_2k_3[\Delta k(z)]^3} \label{eq:delta-b-square}
\end{equation}
with $s_{123}=6,2,1$ for equilateral, isosceles, and scalene triangles respectively,  $k_\mathrm{f}$ represents the fundamental frequency of the survey as was described in the previous section. Further, due to the smallness of the anisotropies, we approximate $P_{21}^\mathrm{D}$ as $P_{21}^0$, \eqref{eq:power_sp_eq}, i.e. the total power spectrum including thermal noise. 
Additionally, for degenerate configurations, i.e.,  $k_1=k_2+k_3$, we multiply the bispectrum variance by a factor of 2 \cite{LSG_bias:2016bnm}. 

For the quadrupolar asymmetry, the Fisher matrix will be written as
\begin{equation}
    F^Q_{B,ij}(z) = \frac{1}{4\pi}\sum_{k_1k_2k_3}\int\d\mu_1\int_0^{2\pi}\d\phi \frac{\partial B_{21}^Q}{\partial g_i} \frac{\partial B_{21}^Q}{\partial g_j} \frac{1}{\Delta B^2}
\end{equation}
with $B_{21}^Q$ given in \eqref{eq:bispectrum_quad} and $\Delta B^2$ in \eqref{eq:delta-b-square} with $P^\mathrm{Q}_{21}$ instead of $P^\mathrm{D}_{21}$. In addition to functions $\mathfrak{F}_i$ having $\phi$ dependencies, the kernels $Z_i$ are also be $\phi$ dependent. {Thus unlike power spectrum case Subsection~\ref{subsec:Power_spectrum_result}, there won't be any identical constraint on the parameters $g_1$ and $g_2$ (and similarly $g_3$ and $g_4$) this time and we need to constrain all 5 parameters.}

\begin{figure}[t]
\centering
\includegraphics[width=0.32\textwidth]{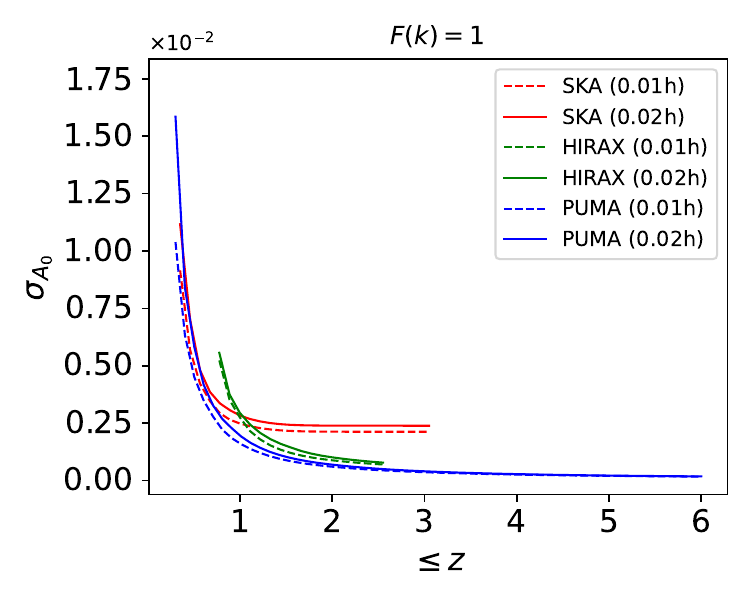}
\includegraphics[width = 0.32\textwidth]{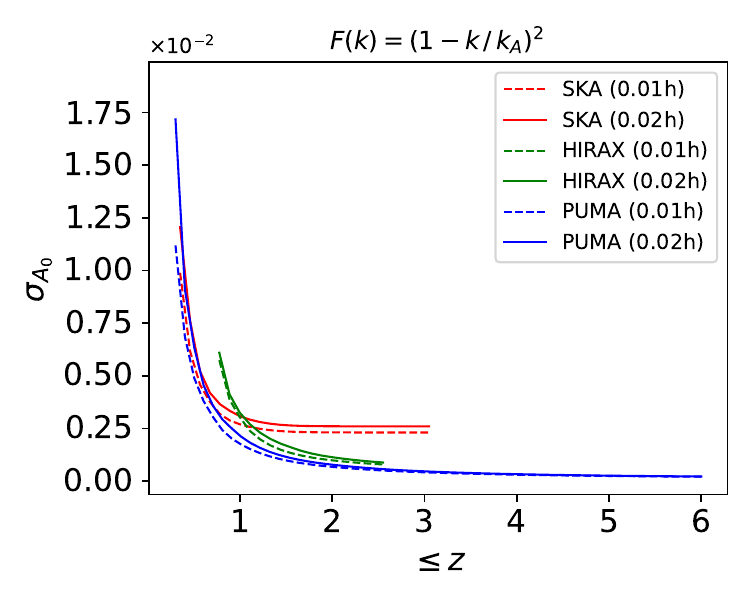}
\includegraphics[width=0.32\textwidth]{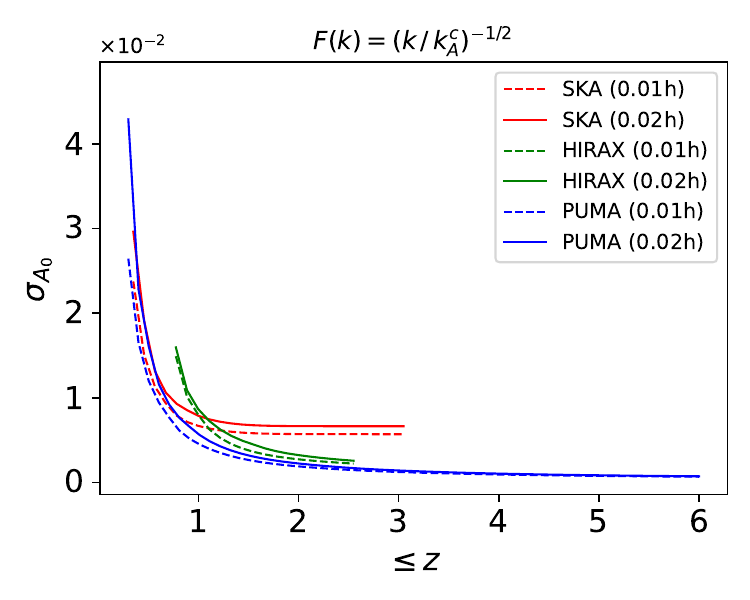}
\caption{As in Fig \ref{fig:Dipolar_Power_z_kmaxFix} but for bispectrum. In this case too, PUMA gives the best constraint in the common redshift range. 
}
\label{fig:dipolar_bispectrum}
\end{figure}

\begin{figure}[t]
\centering
\includegraphics[width=0.32\textwidth]{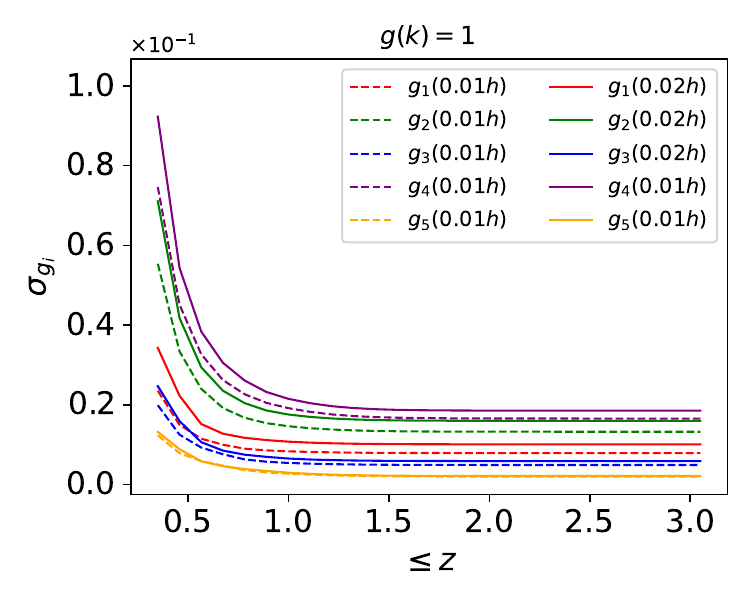}
\includegraphics[width = 0.32\textwidth]{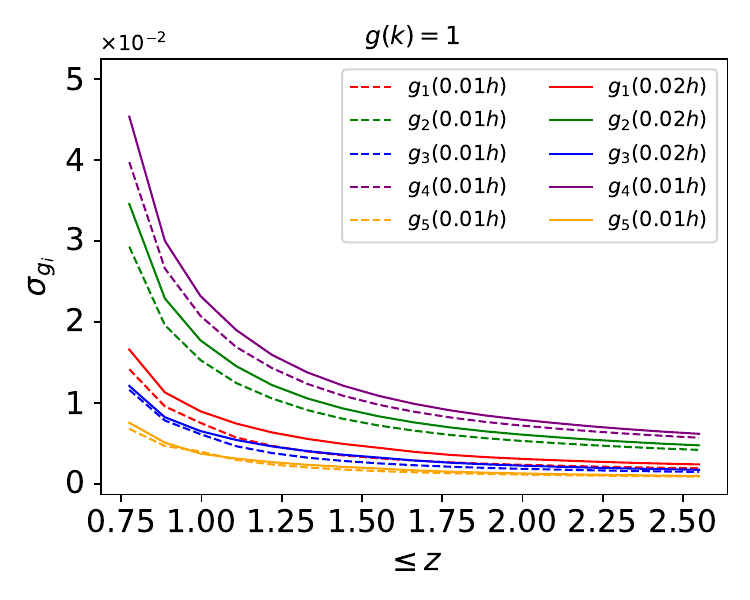}
\includegraphics[width=0.32\textwidth]{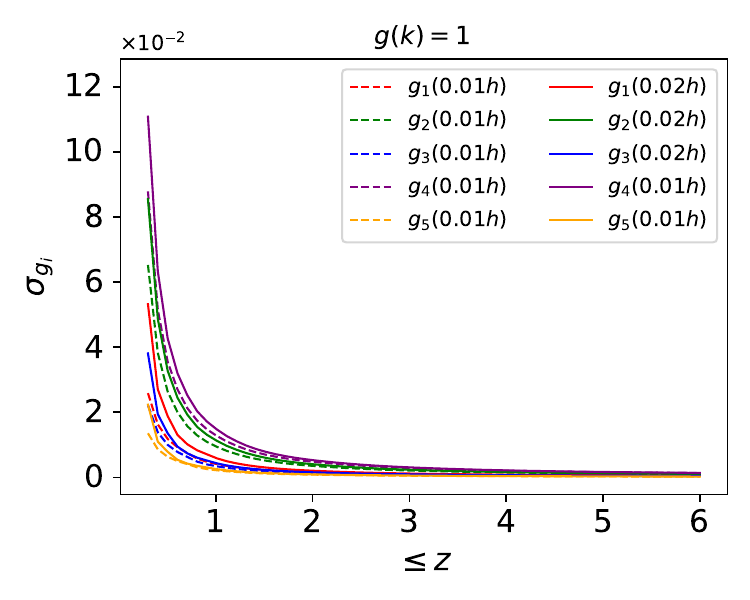}
\caption{Same as Fig \ref{fig:Quadru_Power_g1_all_surveys} but for bispectrum. Again, we find that $g_5$ is the best constrained parameter.}
\label{fig:Quad_Bispec_all_g_i}
\end{figure}

\begin{figure}[t]
\centering
\includegraphics[width=0.4\textwidth]{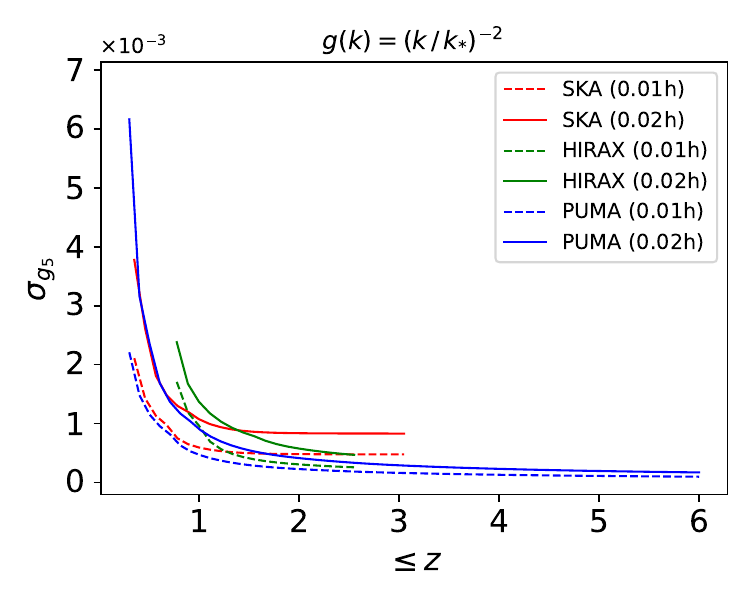}
\includegraphics[width =0.4\textwidth]{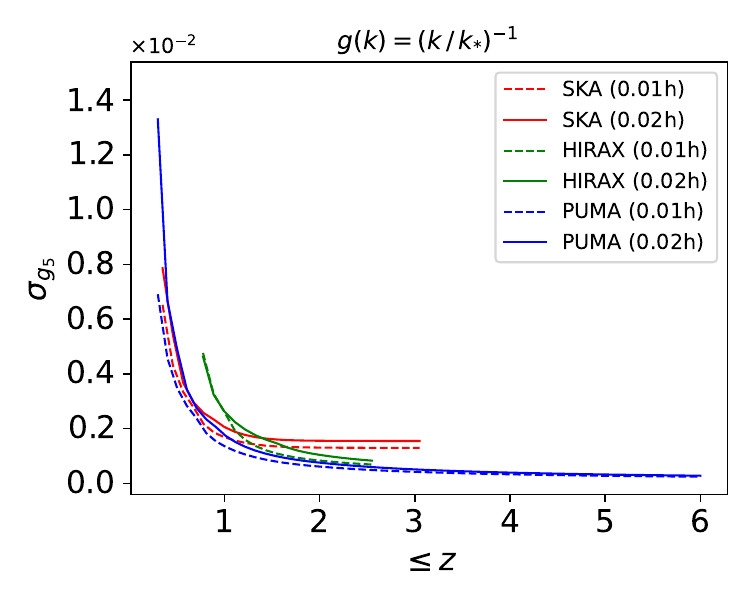} \\
\includegraphics[width=0.4\textwidth]{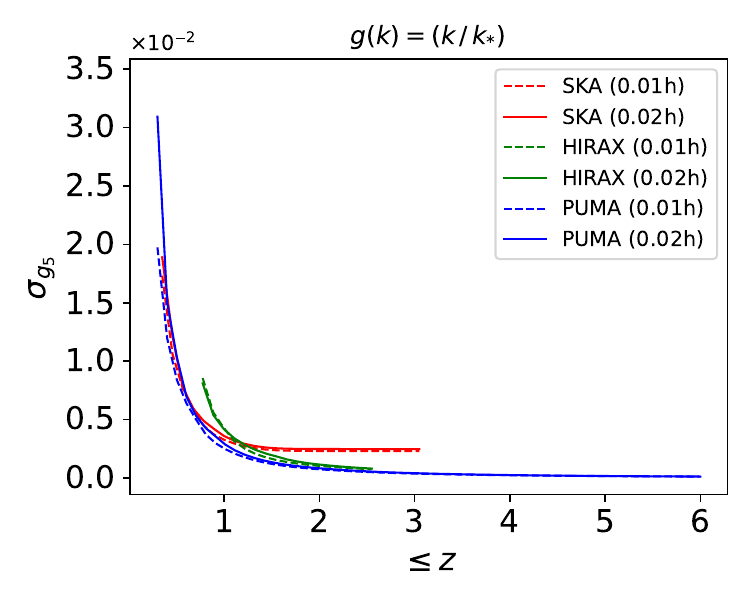}
\includegraphics[width=0.4\textwidth]{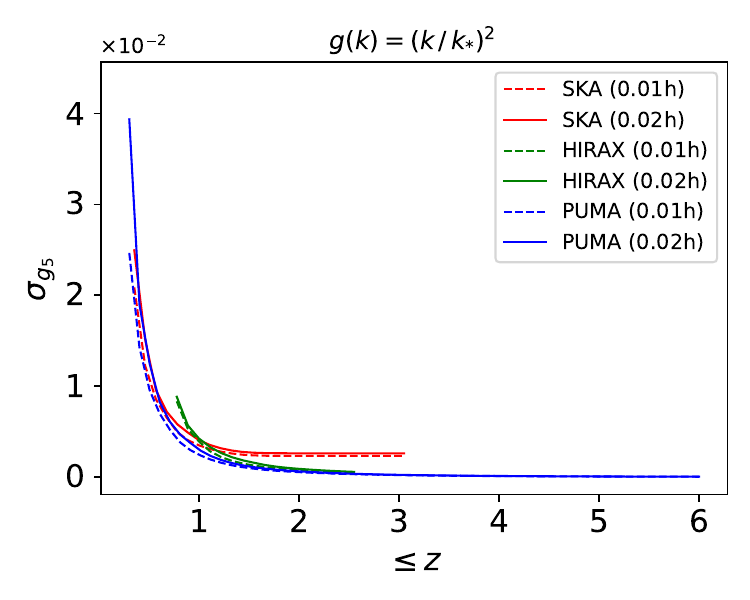}
\caption{Same as in Fig \ref{fig:Quadru_powerZ_kmaxFix} but for bispectrum.}
\label{fig:Quadru_bispec}
\end{figure}

We must point out here that the bispectrum results are sensitive to the chosen values of $\Delta \mu_1$, $\Delta\phi$ \cite{Ivanov:2021kcd} used for integration. Following \cite{Volume:2024tvi}, in this study, we take $\Delta\mu = 0.04$ and $\Delta\phi=\pi/25$. We give the constraints on dipolar anisotropy using bispectrum in Figure \ref{fig:dipolar_bispectrum}, while quadrupolar anisotropy results are given in Figures \ref{fig:Quad_Bispec_all_g_i} and \ref{fig:Quadru_bispec}. As was done in the case of power spectrum, in Figure \ref{fig:Quad_Bispec_all_g_i}, we first check the $\sigma_{g_i}$ trend for $g(k)=1$. We find that just like power spectrum, bispectrum gives best constraints on $g_5$. Thus for the rest of the scalings shown in Figure \ref{fig:Quadru_bispec}, we show cumulative error only for $g_5$. 
We find that the constraints on the anisotropy parameters using the bispectrum are of the same order as that of the power spectrum. Also, similar to the power spectrum, the bispectrum also shows that PUMA has better constraining power. This reflects the conventional understanding that distributing the total collecting area across many small interferometric elements, such as PUMA’s $\sim32,000$ dishes, enhances sensitivity by broadening the field of view.

We also observe that the constraining power of the bispectrum is more sensitive to foregrounds as compared to the power spectrum. In order to quantify this, we define the percent change in the cumulative error for $k_\mathrm{fg} = 0.01h\mathrm{Mpc}^{-1}$ and $k_\mathrm{fg} = 0.02h \mathrm{Mpc}^{-1}$ by
\begin{equation}
    \rho = \frac{\sigma_{0.02h}(z)- \sigma_{0.01h}(z)}{\sigma_{0.02h}(z)} \times 100 \label{eq:rho_defined} 
\end{equation}
where $\sigma_x(z)$ is the cumulative error obtained for redshift $\le z$ with $k_\mathrm{fg}=x$. This quantity, for both dipolar and quadrupolar anisotropies is tabulated in Tables \ref{tab:rho_dipolar} and \ref{tab:rho_quadru} respectively. From Table \ref{tab:rho_dipolar}, it is clear that percent cumulative change $\rho$ for bispectrum for a given $z$ and survey is greater than or equal to that of power spectrum. E.g., for PUMA $\rho$ values obtained for bispectrum is 20\% which means that when the foregrounds are changed from $0.02h$ to $0.01h$, the change is 20\% while for power spectrum the change is merely 6\%.  The same pattern is seen in Table \ref{tab:rho_quadru}.

\begin{table}
\centering
\begin{tabular}{l c c@{\hspace{.5cm}}c c@{\hspace{.5cm}}c c@{\hspace{.5cm}}c}
\hline\hline
$\lesssim z$ & & \multicolumn{2}{c}{SKA} & \multicolumn{2}{c}{HIRAX} & \multicolumn{2}{c}{PUMA} \\
\hline
& & Pow & Bisp. & Pow & Bisp. & Pow & Bisp. \\
\hline
$1$ & & 5 & 16 & 5 & 16  & 6 & 20 \\
$2.5$ & & 4 & 15 & 3 & 12 & 3 & 13 \\
\hline\hline
\end{tabular}
\caption{Percent change in the cumulative error $\rho$, \eqref{eq:rho_defined} for the dipolar anisotropy parameter $A_0$ with the scaling $F(k) = 1$. For comparison sake, we have chosen $z$ values from the common redshift range for all three surveys.}
\label{tab:rho_dipolar}
\end{table}

\begin{table}
\centering
\begin{tabular}{l c c@{\hspace{.5cm}}c c@{\hspace{.5cm}}c c@{\hspace{.5cm}}c}
\hline\hline
$\lesssim z$ & & \multicolumn{2}{c}{SKA} & \multicolumn{2}{c}{HIRAX} & \multicolumn{2}{c}{PUMA} \\
\hline
& & Pow & Bisp. & Pow & Bisp. & Pow & Bisp. \\
\hline
$1$ & & 1 & 9 & 4 & 5  & 4 & 14 \\
$2.5$ & & 1 & 7 & 3 & 8 & 4 & 10 \\
\hline\hline
\end{tabular}
\caption{Same as in Table \ref{tab:rho_dipolar} but for quadrupolar anisotropy parameter $g_5$ with $g(k) = 1$.}\label{tab:rho_quadru}
\end{table}

\section{Conclusion and Outlook \label{eq:Conclu_Outlook}}
In this article, we have discussed the potential of 21cm maps in constraining statistical isotropy, one of the most important symmetries on which modern cosmology is based. We studied both dipolar and quadrupolar asymmetries in the primordial power spectrum of the curvature perturbations using 21cm intensity mapping during post-reionization era using Fisher formalism. We employed both power \& bispectra to constrain both dipolar and quadrupolar anisotropy parameters. Our key findings are:


\begin{enumerate} 
\item For all three surveys, the cumulative error $\sigma$ on both dipolar ($A_0$) and quadrupolar ($g_5$) anisotropy parameters, employing power or bispectrum is $\lesssim 10^{-3}$. In comparison to this, the CMB and galaxy survey power spectrum analyses of quadrupolar anisotropy \cite{Planck:2018jri,Quadrupol_SDSS}, the 1$\sigma$ error on $g_{20}$ is $\Delta g_{20} \approx 10^{-2}$. But from \eqref{eq:Quadru_Asymm_Unsimp} and \eqref{eq:g-def}, $g_5 \approx g_{20}/3$. This means that future 21cm surveys have one order of magnitude better constraining power as compared to CMB or galaxy surveys. 
\item Further, out of all quadrupolar anisotropy parameters, $g_i$, the best constraints are obtained on $g_5$. 
\item Out of all the surveys we have used in the study, PUMA works the best in the common redshift range for both power and bispectrum. 
\item The cumulative percent change in the case of bispectrum, when the foreground cuts are changed is found to be larger as compared to power spectrum. Thus, the constraining power of bispectrum is more sensitive to the foregrounds. This is evident by the percentage change in cumulative error $\rho$ given in Tables \ref{tab:rho_dipolar} and \ref{tab:rho_quadru} for dipolar and quadrupolar parameters. We believe that this is due to the fact that the bispectrum for its description needs information of three Fourier modes in contrast to the power spectrum that needs only two. Thus, a bispectrum configuration depending upon three Fourier modes is more likely to be discarded on account of the foreground conditions as compared to the power spectrum, which needs only two modes. To test our hypothesis, we calculate the change in the number of allowed configurations for both the power spectrum and the bispectrum for HIRAX at $z=1$ when $k_\mathrm{fg}$ is changed from $0.01h\mathrm{Mpc}^{-1}$ to $0.02h\mathrm{Mpc}^{-1}$. The results are shown in Table \ref{tab:hypo_test}. It can be seen that the percent change in the number of allowed configurations for bispectrum is $\sim6$ times as large as that of power spectrum. This continues to be case for other cases for various surveys and redshifts.
\end{enumerate}
\begin{table}
    \centering
    \begin{tabular}{l l l l}
    \hline\hline
     & $0.01h$ & $0.02h$ & \% change \\
    \hline
    Power Spectrum & 240,518 & 224,550 & $-6.6$\\
    Bispectrum & 8,271,008 & 5,166,738 & $-37.5$ \\
    \hline\hline
    \end{tabular}
    \caption{Total number of allowed configurations in power spectrum and bispectrum as a function of foreground cuts $k_\mathrm{fg}$. The numbers are shown for HIRAX at $z=1$.}
    \label{tab:hypo_test}
\end{table}


\acknowledgments
Rahul Kothari acknowledges computing facilities availed through IIT Mandi seed Grant IITM/SG/DIS-ROS-SPA/111. Bhuwan Joshi acknowledges HTRA PhD scholarship while this work was done. We are also very thankful to Roy Maartens, Heyang Long, Sukhdeep Singh Gill, Manoj Rana and Suman Majumdar for discussions. We are also very thankful to the anonymous referee, whose comments helped us considerably improve the quality of the paper. 

\appendix

\section{Calculation of azimuthal and polar angles\label{sec:azimutha-and-polar}}
In this appendix, our objective is to calculate the azimuthal and polar angles of the three vectors $\mathbf{k}_i$ that form the triangle, i.e., $\sum_i \mathbf{k}_i=0$. Although, the expressions for azimuthal angles, i.e., $\mu_2$ and $\mu_3$ is a known result \cite{Fisher_joint:2018uaz}, but to the best of our knowledge there are no such results in the literature for polar angles. So for completeness sake, here we give a complete derivation of calculations of all the involved angles.  We assume a general LoS direction and later specialize to our case by considering it along $\hat{z}$. We want to express a vector $\mathbf{x}$ in terms of known vectors $\bm{\alpha}$ and $\bm{\beta}$ given (a) the angle between $\mathbf{x}$ and $\bm{\alpha}$ is $\theta$ and (b) the angle $\phi\in [0,2\pi]$ between $\mathbf{n}=\bm{\alpha}\times \mathbf{x}$ and the projection of $\bm{\beta}$ in the plane formed by the vectors $\mathbf{n}$ and $\mathbf{n}\times\bm{\alpha}$. Thus, 
\begin{equation}
    \cos\phi = \frac{\alpha}{\sqrt{R}}\left[\bm{\beta} - \frac{(\bm{\beta}\cdot \bm{\rho})\bm{\rho}}{\rho^2}\right]\cdot \frac{\mathbf{n}}{n},\quad R = \alpha^2\beta^2 - (\bm{\alpha}\cdot\bm{\beta})^2 \label{eq:phi-def}
\end{equation}
where $\bm{\rho}$ is normal to the plane formed by $\mathbf{n}=\bm{\alpha}\times \mathbf{x}$ and $\mathbf{n}\times\bm{\alpha}$. Although, there can be two normal vectors, in our case we take
\begin{equation}
    \bm{\rho} = (\mathbf{n}\times\bm{\alpha})\times \mathbf{n}
\end{equation}
In order to express $\mathbf{x}$ in terms of the given vectors and angles $\theta$ and $\phi$, the first step is to write the vector $\mathbf{x}$ in terms of linear combination of right handed orthogonal vector basis, \textit{viz.}, $\{\bm{\alpha}, \bm{\beta}\times \bm{\alpha}, \bm{\alpha}\times (\bm{\beta}\times\bm{\alpha} )\}$
\begin{equation}
    \mathbf{x} = a\bm{\alpha} + b (\bm{\beta}\times \bm{\alpha}) + c (\bm{\alpha}\times (\bm{\beta}\times\bm{\alpha})) \label{eq:x-vector-ortho-triad}
\end{equation}
As the vectors are mutually orthogonal, the basis is linearly independent. 
Our goal now is to calculate these coefficients. The coefficient $a$ is easily calculated by taking dot product of \eqref{eq:x-vector-ortho-triad} with $\bm{\alpha}$ and we find
\begin{equation}
    a = \frac{x\cos\theta}{\alpha}
\end{equation}
To calculate coefficients $b$ and $c$, we first notice that 
\begin{equation}
    n^2 = |\bm{\alpha}\times \mathbf{x}|^2 = x^2 \alpha^2 \sin^2\theta = \alpha^2 \left[ c^2 \alpha^2 |\bm{\alpha}\times \bm{\beta}|^2 + b^2 R \right] \label{eq:b-c-condition1}
\end{equation}
Another equation for these coefficients is obtained from \eqref{eq:phi-def} which after plugging in expressions of $\bm{\rho}$ \& $\mathbf{n}$ gives 
\begin{equation}
    n^2\cos^2\phi= b^2 \alpha^2 R \label{eq:b-c-condition2}
\end{equation}
Now we have two equations \eqref{eq:b-c-condition1} and \eqref{eq:b-c-condition2}, solving these two gives
\begin{equation}
    c = \frac{x\sin\theta \sin\phi}{\alpha\ |\bm{\alpha}\times\bm{\beta}|},\quad b = \frac{x\cos\phi \sin\theta}{\sqrt{R}}
\end{equation}
where we have only considered the positive roots to get results consistent with the literature. Now we can check our results for the particular case when $\bm{\alpha}= \mathbf{k}_1$, $\mathbf{x}=\mathbf{k}_2$ as two sides of the triangle and LoS along Z axis, i.e., $\bm{\beta}=\hat{z}$. Plugging in these and after some simplification we get our coefficients as
\begin{equation}
    a= \frac{k_2\cos\theta_{12}}{k_1},\ b = \frac{k_2\cos\phi \sin\theta_{12}}{k_1\sqrt{1-\mu^2}},\ c = \frac{k_2\sin \theta_{12}\sin\phi}{k_1^2\sqrt{1-\mu^2}}
\end{equation}
where $\mu$ is the angle between $\mathbf{k}_1$ and $\hat{z}$. Taking dot product of \eqref{eq:x-vector-ortho-triad} with $\hat{z}$ thus gives $\mu_2$ which we find to be
\begin{equation}
    \mu_2 = \mu \cos\theta_{12} + \sqrt{1-\mu^2} \sin\theta_{12}\sin\phi
\end{equation}
Using the fact that vectors $\mathbf{k}_i$ form a triangle, we calculate $\mu_3$ using  $\mathbf{k}_3\cdot \hat{z}$. Now we can proceed to calculating the polar angles, to that end, we notice that the polar angle $\Phi$ for a given vector $\mathbf{X}$ is
\begin{equation}
    \tan\Phi = \frac{\mathbf{X}\cdot \hat{y}}{\mathbf{X}\cdot \hat{x}}
\end{equation}
Conventionally $\mathbf{k}_1$ is taken in the XZ plane so that $\phi_1=0$. So we only need to calculate $\phi_2$ and $\phi_3$ which we find 
\begin{align}
    \tan\phi_2 &= \frac{\cos\phi\sin\theta_{12}}{\sqrt{1-\mu^2} \cos\theta_{12} - \mu \sin\theta_{12}\sin\phi},\\ \tan\phi_3 &=  \frac{k_2 \cos\phi \sin\theta_{12}}{\sqrt{1-\mu^2}(k_1 + k_2 \cos\theta_{12}) - \mu k_2 \sin\theta_{12}\sin\phi}
\end{align}

\bibliographystyle{JHEP}
\bibliography{reference}
\end{document}